\documentclass[iop,apj]{emulateapj}
\usepackage{times}
\usepackage{amsmath}

\usepackage[breaklinks,colorlinks,citecolor=blue,linkcolor=magenta]{hyperref} 

\usepackage[all]{hypcap} 

\shorttitle{ELQS in SDSS : Candidate Selection}

\shortauthors{Schindler et al.}

\begin{document}
 
\title{The Extremely Luminous Quasar Survey (ELQS) in the SDSS footprint I.: \\ Infrared Based Candidate Selection} 

\author{Jan-Torge Schindler\altaffilmark{1}}
\author{Xiaohui Fan\altaffilmark{1}}
\author{Ian D. McGreer\altaffilmark{1}}
\author{Qian Yang\altaffilmark{2,3}}
\author{Jin Wu\altaffilmark{2,3}}
\author{Linhua Jiang\altaffilmark{2}}
\author{Richard Green\altaffilmark{1}}

\altaffiltext{1}{Steward Observatory, University of Arizona, 933 North Cherry Avenue, Tucson, AZ 85721,USA}
\altaffiltext{2}{Kavli Institute for Astronomy and Astrophysics, Peking University, Beijing 100871, China}
\altaffiltext{3}{Department of Astronomy, School of Physics, Peking University, Beijing 100871, China}

\begin{abstract}
Studies of the most luminous quasars at high redshift directly probe the evolution of the most massive black holes in the early Universe and their connection to massive galaxy formation. 
However, extremely luminous quasars at high redshift are very rare objects. Only wide area surveys have a chance to constrain their population.
The Sloan Digital Sky Survey (SDSS) has so far provided the most widely adopted measurements of the quasar luminosity function (QLF) at $z>3$. However, a careful re-examination of the SDSS quasar sample revealed that the SDSS quasar selection is in fact missing a significant fraction of $z\gtrsim3$ quasars at the brightest end.
We have identified the purely optical color selection of SDSS, where quasars at these redshifts are strongly contaminated by late-type dwarfs, and the spectroscopic incompleteness of the SDSS footprint as the main reasons.
Therefore we have designed the Extremely Luminous Quasar Survey (ELQS), based on a novel near-infrared JKW2 color cut using WISE AllWISE and 2MASS all-sky photometry, to yield high completeness for very bright ($m_{\rm{i}} < 18.0$) quasars in the redshift range of $3.0\leq z\leq5.0$. It effectively uses random forest machine-learning algorithms on SDSS and WISE photometry for quasar-star classification and photometric redshift estimation.
The ELQS will spectroscopically follow-up $\sim 230$ new quasar candidates in an area of $\sim12000\,\rm{deg}^2$ in the SDSS footprint, to obtain a well-defined and complete quasars sample for an accurate measurement of the bright-end quasar luminosity function at $3.0\leq z\leq5.0$. In this paper we present the quasar selection algorithm and the quasar candidate catalog.

\end{abstract}


\section{Introduction}
Quasars are the most luminous non-transient light sources in the Universe. Powered by the accretion onto super-massive black holes (SMBHs) in the centers of galaxies, they provide important probes for the formation and evolution of structure in the Universe up to the highest redshifts.
As strong light-beacons their emission traverses the intergalactic medium on their way to us and allows to study its properties and evolution. 
Furthermore quasars produce large quantities of ionizing photons that drive the He-reionization of the Universe \citep[e.g.][]{Haiman1998, Madau1999, MiraldaEscude2000}. 
The discovery of luminous quasars $0.8\,\rm{Gyr}$ after the Big Bang \citep{Mortlock2011}, places strong constraints on the formation and growth mechanisms of SMBHs.

A fundamental probe of the growth and evolution of SMBHs over cosmic time is the quasar luminosity function (QLF). It is a measure of the spatial number density of quasars as a function of magnitude (or luminosity) and redshift.
From $z=0$ on the number density of quasars increases \citep{Schmidt1968} up to the peak epoch of quasar activity at redshifts around $z=2-3$. At redshifts beyond $z\approx3$ the quasar number density is found to decline strongly \citep[e.g.][]{Schmidt1995, Fan2001b, Richards2006, Ross2013}.
The QLF is best fit with a broken power law \citep{Boyle1988, Boyle2000, Pei1995} with a steep power law slope at high luminosities and  a flatter power law slope towards lower luminosities. The slopes, the break point and the overall normalization are known to evolve with redshift and may provide insight into the physical mechanism of BH growth across cosmic time and the structure formation of the Universe.
For instance, the ratio of bright quasars to faint quasars decreases from redshift $z\approx4$ to $z\approx1$, an indication that the brighter quasars, associated with the more massive black holes, finish their evolution first \citep{Ueda2003,Marconi2004,Labita2009}.


Investigations into the evolution of the QLF have previously found the bright-end slope to be flattening \citep{Koo1988, Schmidt1995, Fan2001b, Richards2006}. However more recent measurements have now established that it remains steep up to redshifts of $z\sim6$ \citep{Jiang2008, Croom2009, Willott2010a, McGreer2013, Yang2016}.

The difficulty in measuring the bright-end QLF is partly due to the rapid decrease in spatial density of quasars towards high luminosities and the overall decline of their number density towards higher redshifts. Therefore only wide area surveys allow for reliable measurements of the bright-end slope. 

In addition, purely optical color selections are biased against certain redshift ranges \citep{Richards2006, Ross2013}. While this bias was well accounted for in previous calculations of the quasar luminosity function, the low number of quasars at the brightest end does not allow for secure statistics.

Including its latest phase the Sloan Digital Sky Survey (SDSS; \citet{York2000}) has covered $14,555\,\rm{deg}^2$ in the northern hemisphere with five band (\textit{ugriz}) optical imaging and extensive spectroscopic follow-up. 
The first phase of the survey allowed for the discovery of the first $z\geq5$ quasar \citep{Fan1999b}. The quasar surveys of the first and second phase of SDSS led to the DR7 quasar catalog (DR7Q; \citep{Schneider2010}) which contains $\geq105,000$ quasars. 

The SDSS-III Baryon Oscillation Spectroscopic Survey (BOSS; \citet{Eisenstein2011, Dawson2013}) and the SDSS-IV extended Baryon Oscillation Spectroscopic Survey (eBOSS; \citet{Dawson2016}) have carried out dedicated spectroscopic follow-up of galaxies and quasars up to $z\approx3$. The DR14 quasar catalog (DR14Q; Paris, I. et al. in preparation, see also \citet{Paris2017} for DR12Q), the largest quasar sample to date, now includes more than $500,000$ known quasars.

However, we have discovered that the SDSS quasar surveys have missed many bright (SDSS $m_{\rm{i}}<18.5$) higher redshift ($3.0<z<5.0$) quasars. 
The optically based color selection of quasar candidates in SDSS, BOSS and eBOSS has relatively low completeness in redshift regions, where the stellar locus overlaps with quasars in optical color space \citep{Richards2006, Ross2013}. Furthermore $z>3$ quasars are not explicitly targeted in BOSS and eBOSS and the spectroscopic follow-up of SDSS-I/II is not fully complete, especially in the fall sky ($\rm{RA} {>} 270\,\rm{deg}$ and $\rm{RA} {<} 90\,\rm{deg}$) of the SDSS footprint.
As a result a substantial fraction of bright $3.0<z<5.0$ quasars have been missed in previous estimations of the QLF. 

The power of near- to mid-infrared photometry to comprehensively select quasars that are otherwise indistinguishable from stars in optical bands was exploited with the advent of large infrared surveys.
New quasar selection methods were developed using the infrared K-band excess in the UKIRT (UK Infrared Telescope) Infrared Deep Sky Survey (UKIDSS; \citet{Warren2000,Hewett2006,Maddox2008}), to efficiently separate quasars and stars at lower \citep[$z<3$][]{Chiu2007} and higher \citep[$z>6$][]{Hewett2006} redshifts. 
A range of efforts combined optical and near-infrared photometry. \citet{Barkhouse2001} used the Two Micron All Sky Survey  \citep[2MASS][]{Skrutskie2006} and optical photometry from the Veron-Cetty \& Veron catalog \citep{Veron2000}, whereas \citet{Wu2010} and \citet{Wu2011} combined SDSS and UKIDSS photometry.
In the mid-infrared the Wide-field Infrared Survey Explorer mission (WISE; \citet{Wright2010}) provided deep photometry to further increase the efficiency of $z<3.2$ quasar selections\citet{Wu2012}.

We designed the Extremely Luminous Quasar Survey (ELQS) to re-examine the SDSS footprint. This paper (Paper I) motivates the survey and outlines our candidate selection. A follow-up paper (Paper II) will contain the spectroscopic results of ELQS spring sample along with a first estimate of the bright-end QLF.
At completion of the survey we will summarize all spectroscopic discoveries, calculate the bright-end QLF over the entire surveyed footprint ($\sim12000\,\rm{deg}^2$) and discuss the resulting implications (Paper III).

We first describe the photometric data used for our candidates selection (Section\,\ref{sec_photometry}). Thereafter we carefully analyze why previous SDSS surveys missed bright higher redshift quasars (Section\,\ref{sec_missed_qsos}). We further develop a solution to the incomplete selection via near-infrared/infrared photometry and discuss rejection of extended objects in  Section\,\ref{sec_color_cut}. 
We continue to describe in detail how we employ machine learning algorithms to classify quasar candidates and obtain redshift estimates in Section\,\ref{sec_rf}.
At last we present the construction of the ELQS quasar candidate catalog (Section\,\ref{sec_qso_catalog}) and then summarize the results of this paper in Section\,\ref{sec_conclusion}.

All magnitudes are displayed in the AB system \citep{Oke1983} and corrected for galactic extinction \citep{Schlafly2011}, unless otherwise noted.
We adopt the standard flat $\Lambda\rm{CDM}$ cosmology with $H_0=70\,\rm{km}\rm{s}^{-1}\rm{Mpc}^{-1}$, $\Omega_{\rm{m}}=0.3$ and $\Omega_\Lambda=0.7$, generally consistent with recent measurements \citep{PlanckCollaboration2016}.

\section{Photometry}\label{sec_photometry}
For our quasar candidate selection we use a combination of near-IR 2MASS and mid-IR WISE photometry, complemented with optical photometry from SDSS.

\subsection{The Sloan Digital Sky Survey (SDSS)}
For all SDSS sources we use the point spread function asinh magnitudes \citep{Lupton1999} in the five optical band passes (\textit{ugriz}) \citep{Fukugita1996}. Throughout this paper we only use AB magnitudes. We therefore converted the SDSS u-band and z-band magnitudes with $u_{\rm{AB}} = u'-0.04\,\rm{mag}$ and $z_{\rm{AB}} = z'+0.02\,\rm{mag}$. 
All magnitudes are corrected for galactic extinction using the extinction values from the Casjobs Data Release 13 (DR13) PhotoObjAll or PhotoPrimary tables \citep{Schlafly2011}.
The imaging data of the SDSS survey was completed in 2009 and covers a unique area of $14,555\,\rm{deg}^2$.
The magnitude limits ($95\%$ completeness for point sources) in the five optical bands are $21.6, 22.2, 22.2, 21.3, 20.7$ for u,g,r,i,z, respectively.

\subsection{The Wide-field Infrared Survey Explorer (WISE)}
WISE mapped the entire sky at 3.4, 4.6, 12, and 22$\,\mu\rm{m}$ (W1, W2, W3, W4). The recent AllWISE data release combines the data from the cryogenic and post cryogenic \citep{Mainzer2011} phases of the mission\footnote{\url{http://irsa.ipac.caltech.edu/cgi-bin/Gator/nph-scan?submit=Select&projshort=WISE}}.
The AllWISE source catalog  achieved $95\%$ photometric completeness for all sources with limiting magnitudes brighter than $19.8$, $19.0$ (Vega: $17.1$, $15.7$), in W1 and W2. Saturation affects sources brighter than $8, 7$ in the W1 and W2 bands.
We restrict ourselves to photometry of the W1 ($3.4\,\mu{\rm{m}}$) and W2 ($4.6\,\mu{\rm{m}}$) infrared bands and convert them to AB magnitudes using $W1_{\rm{AB}} = W1 + 2.699$ and $W2_{\rm{AB}} = W2 + 3.339$.
The WISE photometry is then extinction corrected using the extinction coefficients $A_{\rm{W1}}, A_{\rm{W2}} = 0.189, 0.146$ with the extinction values from the SDSS photometry.

\subsection{The Two Micron All Sky Survey (2MASS)}
2MASS has mapped the entire sky in the near-infrared bands J ($1.25\,\mu{\rm{m}}$), H ($1.65\,\mu{\rm{m}}$) and $\rm{K}_s$ ($2.17\,\mu{\rm{m}}$). 
We use the 2MASS point source catalog (PSC), which was prematched to the WISE AllWise source catalog. The 2MASS PSC is generally complete at a level of $10\sigma$ photometric sensitivity for all sources brighter than $16.7$, $16.4$,  $16.1$ (Vega: $15.8, 15.0, 14.3$) in the J,H and $\rm{K}_s$ bands, respectively. Yet, the 2MASS PSC includes all sources with at least a signal-to-noise ratio of $\rm{SNR}\geq7$ in one band or $\rm{SNR}\geq5$ detections in all three bands. Furthermore due to confusion of sources close to the galactic plane the photometric sensitivity is a strong function of galactic latitude. 
Based on the online documentation\footnote{Figure\,7 on \url{https://www.ipac.caltech.edu/2mass/releases/allsky/doc/sec2_2.html}} we estimate the limiting magnitudes of the 2MASS PSC for higher latitudes to be $\rm{J}=17.7$, $\rm{H}=17.5$, $\rm{K}_s=17.1$.
We generally convert the 2MASS magnitudes to AB using $J_{\rm{AB}} {=} J + 0.894$, $H_{\rm{AB}} {=} H + 1.374$, ${K_s}_{,\rm{AB}} {=} K_s + 1.84$ and correct for galactic extinction ($A_{\rm{J}}, A_{\rm{H}}, A_{\rm{K_s}} {=}  0.723, 0.460, 0.310$).
While we test the machine learning methods using 2MASS photometry, the final selection of the quasar candidate catalog only uses the 2MASS bands for the JKW2 color cut.

\section{Bright High-z Quasars Missed By SDSS }\label{sec_missed_qsos}

\begin{figure*}[t!]
 \centering
 \includegraphics[width=0.95\textwidth]{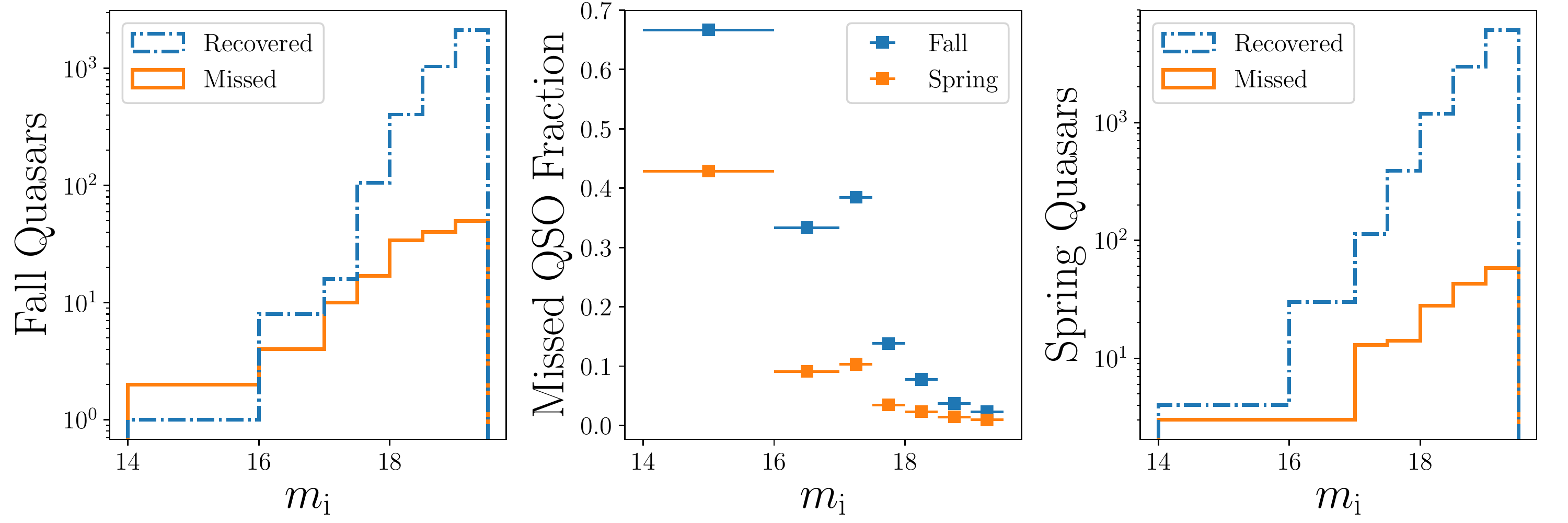}
 \caption{In this Figure we examine the fraction of quasars known in the MQC but \textit{missed} by SDSS at redshifts of $2.5{\leq}z{<}5$. Quasars that were identified by SDSS are termed \textit{recovered}. \textbf{Left:} The number distributions of missed and recovered quasars as a function of i-band magnitude bins in the SDSS \textit{fall} footprint. \textbf{Middle:} The fractions of missed quasars to the total number of quasars in the MQC as a function of i-band magnitude bins, divided into a \textit{fall} sky and \textit{spring} sky sample. \textbf{Right:}  The number distributions of missed and recovered quasars as a function of i-band magnitude bins in the SDSS \textit{spring} footprint.
 This figure illustrates, that the majority of missed quasars are bright ($m_{\rm{i}}{<}17.5$) and located in the \textit{fall} sky of the SDSS footprint.}
 \label{fig_missed_fraction}
\end{figure*}

\begin{table*}[t]
\centering
\caption{Number counts of quasars missed in SDSS to the total number of known quasars in the SDSS matched MQC}
\label{tab_missed_fraction}
\begin{tabular}{c|ccccccc}
\tableline
\tableline
 & \multicolumn{6}{c}{i-band magnitude bins}\\
\tableline
& $14.0 \leq m_{\rm{i}}$ & $16.0 \leq m_{\rm{i}} $ & $17.0 \leq m_{\rm{i}}  $ & $17.5 \leq m_{\rm{i}} $ & $18.0 \leq m_{\rm{i}} $ & $18.5 \leq m_{\rm{i}}  $ & $19.0 \leq m_{\rm{i}}  $ \\
& $< 16.0$ & $< 17.0$ & $< 17.5$ & $< 18.0$ & $< 18.5$ & $< 19.0$ & $< 19.5$ \\
\tableline
\textbf{Fall sky and $\mathbf{2.5\leq z<5}$} & & & & & & \\
\tableline
Missed QSOs&2 & 4 & 10 & 17 & 34 & 40 & 50 \\
Known QSOs& 3 & 12 & 26 & 123 & 438 & 1080 & 2180 \\
Fraction& 0.67 & 0.44 & 0.31 & 0.15 & 0.07 & 0.03 & 0.02 \\
\tableline
\textbf{Spring sky and $\mathbf{2.5\leq z<5}$} & & & & & & \\
\tableline
Missed QSOs&3 & 3 & 13 & 14 & 28 & 43 & 58 \\
Known QSOs& 7 & 33 & 126 & 404 & 1215 & 3014 & 6152 \\
Fraction&0.43 & 0.09 & 0.10 & 0.03 & 0.02 & 0.01 & 0.01 \\
\tableline
\end{tabular}
\end{table*}

\begin{figure*}[t!]
 \centering
 \includegraphics[width=0.9\textwidth]{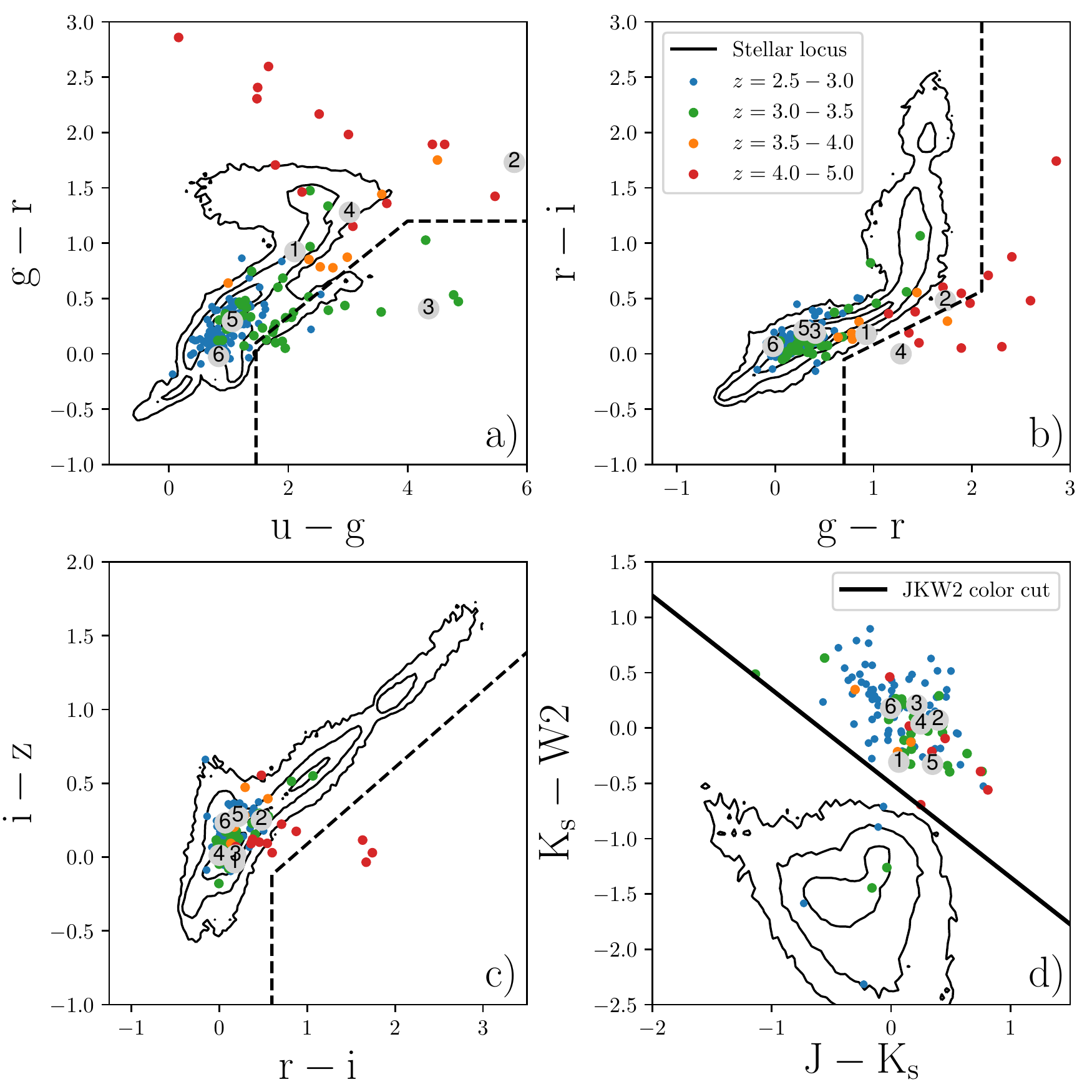}
 \caption{We show the quasars with full SDSS, 2MASS PSC and AllWISE photometry that SDSS missed as filled circles, colored by redshift in the SDSS optical color space (AB magnitudes) in panel a)-c). The numbers reference the missed quasars described in Sec.\,\ref{sec_missed_qso_notes}. A significant fraction of the missed quasars are always overlapping with the stellar locus (black contours). We further display the SDSS high redshift inclusion boxes (ugr in a), $z\geq3.0$; gri in b), $z\geq3.6$; riz in c), $z\geq4.5$). Panel d) shows the population of missed quasars in the near-IR J-Ks and Ks-W2 color-space of 2MASS and WISE. The quasars are well separated from the stellar locus. We display our new JKW2 color cut as the thick black line.}
 \label{fig_cc_missed}
\end{figure*}

In order to investigate why and to what extent SDSS missed bright high redshift quasars, we match a compilation of known quasars in the literature, the Million Quasar catalog \citep[hereafter MQC, Version 5.2][]{Flesch2015}) with SDSS DR14 photometry and the SDSS DR7 and DR14 quasar catalogs.
The MQC includes spectroscopically confirmed quasars as well as high probability quasar candidates from various selection methods. For this exercise we will exclude all candidates.

We restrict ourselves to brighter quasars with $m_{\rm{i}} {<} 19.5$ and at $2.5{\leq}z{<}5$. The sample is further divided into a \textit{spring} ($90\,\rm{deg} {<} \rm{RA} {<} 270\,\rm{deg}$) and \textit{fall} ($\rm{RA} {>} 270\,\rm{deg}$ and $\rm{RA} {<} 90\,\rm{deg}$) portion of the SDSS footprint.
All quasars that have SDSS photometry, but are not included in the DR7 and DR14 quasar catalogs, will be termed \textit{missed} quasars.
We investigate the fraction of missed quasars to all quasars found in MQ in Table\,\ref{tab_missed_fraction} and Fig.\,\ref{fig_missed_fraction}.
Both clearly illustrate that the fraction of missed quasars is larger in the fall sky, than in the spring sky. For $2.5{\leq}z{<}5$ quasars with i-band magnitudes $m_{\rm{i}}{<}17.0, 17.5, 18.5$ the fraction of missed to known quasars is $f\approx 0.40, 0.39, 0.11$ in the fall and $f\approx 0.15, 0.11, 0.03$ in the spring sample.
These numbers demonstrate the poorer spectroscopic completeness of the SDSS survey in the fall sky. Furthermore the fractions increase as we restrict ourselves to brighter i-band magnitudes, emphasizing that especially bright quasars were missed.

The majority (${>}90\%$) of the SDSS and BOSS quasars at $2.5{\leq}z{<}5$  were observed during the BOSS campaign. In the SDSS and BOSS five-band optical color-space quasars at $z=2.5-3.5$ overlap significantly with the stellar locus. The survey completeness of SDSS \citep[][their Figure\,6]{Richards2006} and BOSS \citep[][their Figure\,6]{Ross2013} show the low completeness in those regions.
Furthermore as the candidates get brighter the contamination of the stellar population increases demonstrated by a steep decrease in completeness towards brighter magnitudes in \citet[][their Figure\,6]{Ross2013}.

To illustrate the confusion between quasars at $z=2.5-3.5$ and stars, we show the ugriz color-space diagrams in Figure\,\ref{fig_cc_missed}.
The stellar locus is shown in black contours, while we display the missed quasars with SDSS, 2MASS PSC and AllWISE photometry as filled circles color-coded by redshift. It should be noted that this is a subset of all missed quasars, because some of them are not detected by 2MASS and WISE.
In panels a)-c) the majority of the missed quasars overlaps significantly with the stellar locus. The SDSS inclusion regions for higher redshift quasars \citep{Richards2002} also miss some of the $z>3.5$ quasars that scatter into the stellar locus.

Fiber collisions during observations and quality criteria on the photometry may have contributed to the number of missed quasars.
In addition, quasar lenses can have extended morphologies and will have been likely rejected. In some cases objects with extremely bright apparent magnitudes ($m_{\rm{i}} \lesssim 15.5$) were excluded for spectroscopic follow-up to avoid scattered light from nearby fibers.
However, the main reasons SDSS and BOSS missed bright high redshift quasars are the ugriz optical based candidate selection and the spectroscopic incompleteness of the fall sky footprint.

\subsection{Notes on the Brightest, Missed High Redshift Quasars in the SDSS Spring Sky Footprint}\label{sec_missed_qso_notes}
We have selected all missed quasars with $m_{\rm{i}}<17.0$ and $2.5\leq z<5$ that fall into the well surveyed spring sky of the SDSS footprint to understand why they were missed.
We analyze whether the non-fatal and fatal image quality flags and the high redshift inclusion regions of the original selection \citep{Richards2002} play a role in them being overlooked.
All of the quasars below with $z\geq2.8$ are selected with our selection method and will be included in the ELQS spring sample (Paper II).

\subsubsection*{(1) Q 1208+1011}
This object is a known quasar lens \citep{Magain1992,Bahcall1992} ($m_{\rm{i}}=16.77$) with a redshift of $z=3.80$. One of the non-fatal quality flags is raised and it is not included in any of the high redshift inclusion regions. Finally, this Quasar is located too close to the stellar locus in optical color space (see Fig.\,\ref{fig_cc_missed}), to be selected by previous SDSS campaigns.

\subsubsection*{(2) APM 08279+5255}
This object is a well known, lensed broad absorption line quasar with multiple images at $z=3.91$ \citep{Ibata1999}. It has an SDSS i-band magnitude of $m_{\rm{i}}=14.84$. While it does not show any fatal or non-fatal flags, it is also not selected in the high redshift inclusion regions. The colors of this quasar (see Fig.\,\ref{fig_cc_missed}) place it well away from the stellar locus in ug-gr color space, but it is too bright for the SDSS spectroscopic target list to avoid scattered light from nearby fibers.

\subsubsection*{(3) SDSS J1622+0702A}
This object is a bright ($m_{\rm{i}}=16.87$) binary quasar at $z=3.26$ \citep{Hennawi2010}. The fatal and non-fatal flags are not raised and it actually falls into the UGR inclusion region. This quasar has the correct target flag in SDSS, but it was not observed, because of fiber collisions with a nearby galaxy.

\subsubsection*{(4) B 1422+231}
This lensed quasar has a total of four components and is measured to be at $z=3.62$ \citep{Patnaik1992}. It is very bright with $m_{\rm{i}}=15.31$. While it's image quality flags are not raised and it actually falls into the GRI inclusion region, it is marked as extended (\texttt{type}$=3$) by SDSS. However, it would not have been rejected by our selection for it's morphology (see Sec.\,\ref{sec_gal_rejection}).

\subsubsection*{(5) CSO 167}
CSO 167 is a $z=2.56$ quasar \citep{Sanduleak1984,Everett1995} with $m_{\rm{i}}=16.50$. None of the fatal or non-fatal image quality flags are raised. It is not expected to and also does not fall into one of the high redshift inclusion regions. This quasar is too close to the stellar locus (see Fig.\,\ref{fig_cc_missed}) to be selected in previous SDSS campaigns.

\subsubsection*{(6) CSO 1061}
Similar to CSO167, CSO 1061\citep{Sanduleak1989} is at a lower redshift ($z=2.67$; $m_{\rm{i}}=16.33$) and therefore is not expected to be selected by any of the inclusion regions. None of the fatal or non-fatal image quality flags are raised. This object was also not observed because it is too close to the stellar locus.

\section{Bright Quasar selection based on near-infrared photometry}\label{sec_color_cut}

\begin{figure}
 \includegraphics[width=0.5\textwidth]{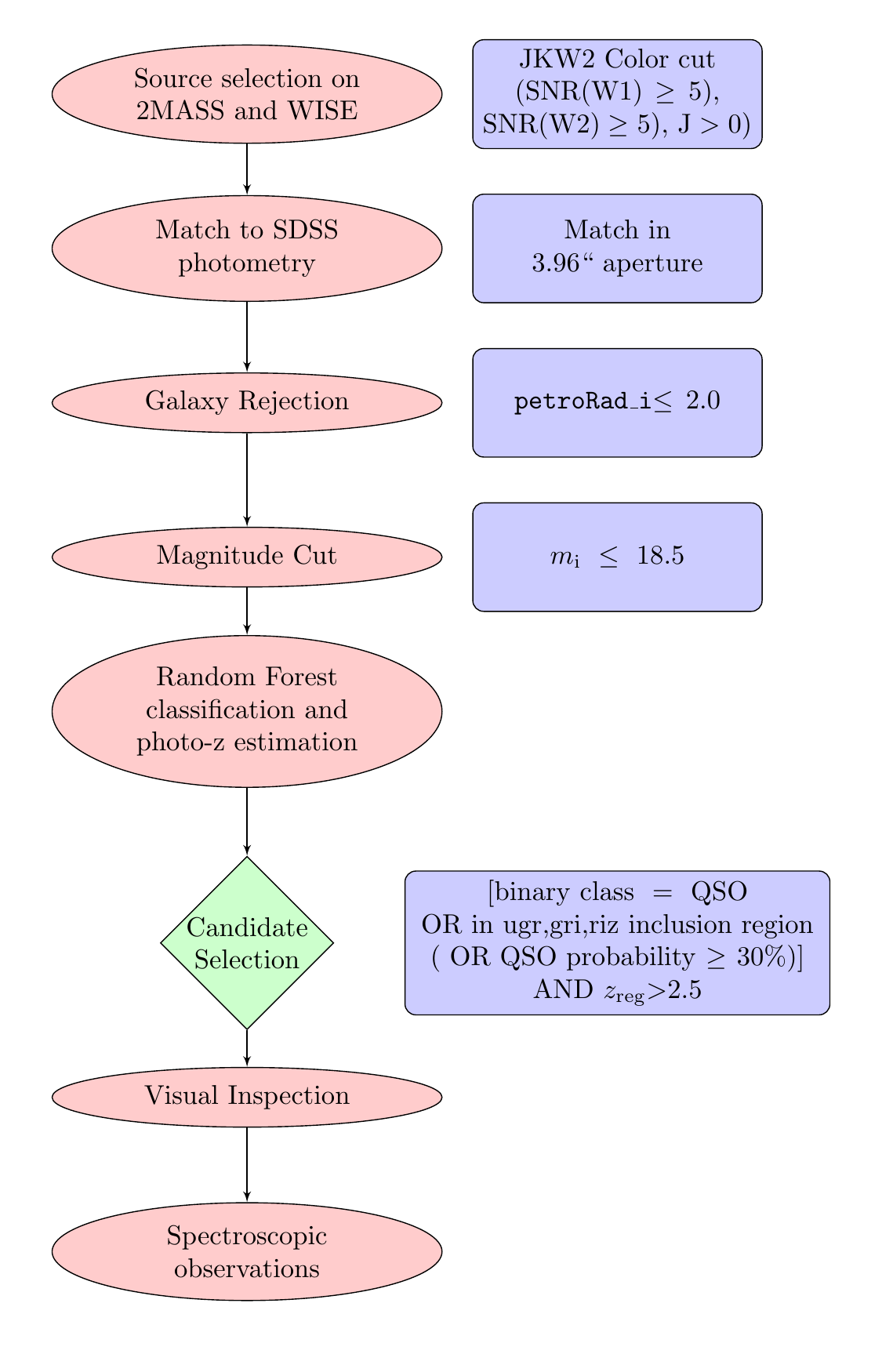}
 \caption{The general steps of the ELQS quasar selection}
 \label{fig_selection_flowchart}
\end{figure}

Our new survey for extremely luminous quasars, ELQS, is designed to bypass the limitations of a purely optical quasar candidate selection by harnessing the information of near-infrared and infrared photometry. For this purpose we have examined the distribution of stars and quasars in the color space of the 2MASS and WISE all-sky surveys.
We discovered that the combination of J-K and Ks-W2 colors offers a clear separation of stars and quasars and designed a JKW2 color cut in this color space.

This color cut is very efficient in rejecting stars and in concert with a measure to eliminate galaxies, the quasar fraction should be fairly pure. Nevertheless, the fraction of bright stars, which make the cut, to bright quasars is non-negligible and the color cut does not discriminate between quasars of different redshifts. In fact, the majority of quasar candidates will be at $z<2.5$ and therefore also reduce the efficiency of our selection. 
Hence, it is necessary to estimate photometric redshifts and to classify candidates at a later stage (Sec.\,\ref{sec_rf}) . 

For an early overview, we illustrate the general steps of our quasar candidate selection in Fig.\,\ref{fig_selection_flowchart}. They will be described in detail in Section\,\ref{sec_qso_catalog}.

\subsection{The JKW2 Color Cut}

While we have shown the overlap between the missed quasars and the stellar locus in optical color-space in Figure\,\ref{fig_cc_missed} panel a)-c) , we also show the 2MASS and WISE J-Ks,Ks-W2 color space in panel d).
Here the stellar locus clearly separates from the distribution of missed quasars. 
After examining the distribution of known quasars to known stars in this color-color space we qualitatively determined the color cut, that separates these distributions best.
In Vega magnitudes the color cut reads,
\begin{equation}
 \rm{Ks}_{\rm{Vega}}-\rm{W2}_{\rm{Vega}} \ge 1.8 - 0.848 \cdot \left(\rm{J}_{\rm{Vega}}-\rm{Ks}_{\rm{Vega}} \right) \ ,
\end{equation}
while in AB magnitudes it changes to
\begin{equation}
 \rm{Ks}-\rm{W2} \ge -0.501 - 0.848 \cdot \left(\rm{J}-\rm{Ks} \right) \ .
\end{equation}

The color cut separates quasars from stars because of their difference in the K-W2 flux ratio (or K-W2 color). This flux ratio is fundamentally different in stars (up to spectral type T) and quasars (up to redshifts of $z\approx5$). For those stars and quasars the stellar flux declines more strongly between the K-band and the W2-band than the quasar flux. Hence, quasars will have a redder K-W2 flux ratio (or color).
Emission lines, like $\rm{H}\alpha$ in the K-band at $z\approx2.1-2.50$ or $\rm{H}\beta$ in the K-band at $z\approx3.2-3.8$, do affect the K-W2 flux ratio, but their influence in negligible.
As a result the JKW2 color cut cannot discriminate between quasars at different redshifts.

Six known quasars lief beneath the JKW2 color cut within the stellar locus in Figure\,\ref{fig_cc_missed} panel d). For none of these objects spectra were available in the literature. A closer examination of their identifications and photometry concluded, that either the photometry or the identification reference seem to be unreliable.

\begin{figure}[t]
 \includegraphics[width=0.45\textwidth]{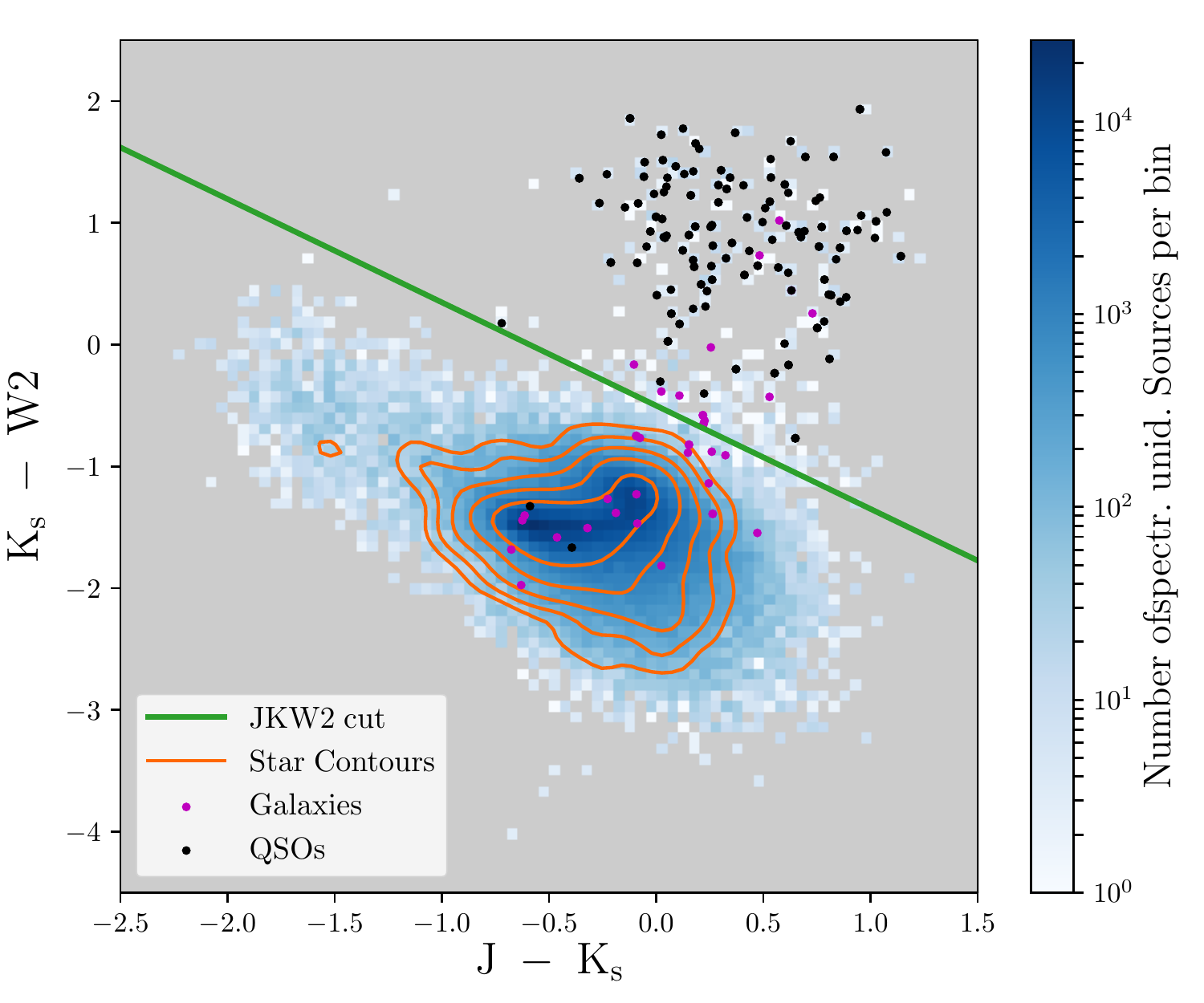}
 \caption{We show all sources with SDSS $m_{\rm{i}}\le18.5$ and \texttt{petroRad\_i}$\le2.0\,\rm{arcsec}$ of a $70\,\rm{deg}^2$ region ($\rm{RA}=120-130$, $\rm{Dec.}=40-50$) in $J$-$K_s$, $K_s$-$W2$ color space. Only sources with full 2MASS and WISE W1 and W2 photometry are selected.
 The green line is the JKW2 color cut and quasars, black dots, lie clearly above the color cut. The majority of spectroscopically unidentified sources, as shown by the white to blue color map, lies below the cut.
 The stellar locus shown by the orange contours coincides with the dark regions of the color map below the cut.
 Spectroscopically identified galaxies are shown as purple dots and straddle the color cut on both sides.
 }
 \label{fig_testfield_colorcut}
\end{figure}

We test the color cut in a $\sim70\,\rm{deg}^2$ region ($\rm{RA}{=}120{-}130\,\rm{deg}$, $\rm{Decl.}{=}40{-}50\,\rm{deg}$), shown in Figure\,\ref{fig_testfield_colorcut}. For this purpose we have selected all sources in SDSS DR13 (PhotoObjAll) brighter than SDSS $m_{\rm{i}} {=} 18.5$, matched the sources with the AllWISE source catalog, including matched 2MASS Point Source Catalog (PSC) photometry, and retrieved spectral identification from the SDSS DR13 (SpecObj) catalog, where possible.

\begin{table*}[t]
\centering
\caption{Selection criterion on $70\,\rm{deg}^2$ test region: JKW2 color cut}
\label{tab_colorcut}
\begin{tabular}{cccccc}
\tableline
\tableline
Data Sample & Quasars & Stars & Galaxies & Full sample & No spectral Id. \\
\tableline
full SDSS colors & 526 & 3,327 & 5,685 & 1,327,439 & 1,317,901 \\
\tableline
SDSS+2MASS+W1W2  & 301 & 2,138 & 5,517 & 1,155,203 & 1,147,247  \\
JKW2 color cut & 298 & 8 & 2,706 & 25,470 & 22,458 \\ 
Fraction & 99.00\% & 0.37\% & 49.05\% & 2.20\% & 1.96\% \\
\tableline
SDSS+2MASS+W1W2  & 207 & 2,081 & 39 & 841,926 & 839,599 \\
\texttt{petroRad\_i}$\le 2.0\,\rm{arcsec}$ &  & & & & \\
JKW2 color cut & 205 & 7 & 15 & 1,296 & 1,069 \\ 
\texttt{petroRad\_i}$\le 2.0\,\rm{arcsec}$ &  & & & & \\
Fraction & 99.03\% & 0.34\% & 38.46\% & 0.15\% & 0.13\% \\
\tableline
\end{tabular}

\end{table*}

The full test field includes a total of 1,327,439 sources detected with full SDSS photometry of which 1,155,203 sources have full 2MASS and WISE W1 and W2 photometry.

For our purposes we estimate the efficiency of the color cut using the knowledge about the spectroscopically identified quasars (at all redshifts), stars and galaxies and the total number of sources with 2MASS and WISE W1 and W2 colors. The results are summarized in Table\,\ref{tab_colorcut}.

The color cut clearly separates stars and quasars. 
It reduces the total number of sources in the test region to 25,470 out of 1,155,203  ($2.20\,\%$). 
The stellar contamination is greatly reduced. Only $0.37\,\%$ (8/2138) of spectroscopically identified stars in SDSS DR13 make the color cut.
Of all 526 quasars in this region 301 have 2MASS and WISE W1 and W2 colors and of these 298 are included in the JKW2 color cut.
Only considering the quasars with full near-infrared photometry, the color cut has an estimated completeness of $\sim99\%$.

The quasar sample in the $70\,\rm{deg}^2$ test region is rather small. Therefore we test our color cut with all quasars from the DR7 and DR12 quasar catalogs that have 2MASS photometry in the J and Ks bands and WISE W2 photometry. 
This sample includes 5945 quasars out of which 5927 are included in the color cut, a fraction of more than $99\%$.

Galaxies, however straddle the JKW2 color cut and only half of all spectroscopically identified galaxies (2706/5517) are excluded. Therefore we believe the majority of the remaining 25470 sources above the color cut to be galaxies.

In total the JKW2 color cut manages to eliminate the majority of stellar sources, while retaining a highly complete sample of bright quasars, uniform in redshift and SDSS i-band magnitude. 
However, in order to efficiently use the color cut we need to first reject galaxy contaminants and secondly estimate the photometric quasar redshifts.

\subsection{Photometric completeness of 2MASS and WISE}
The 2MASS and WISE surveys do not reach the depth of optical surveys, like SDSS.
Therefore fainter sources might only show spurious detections or not detections at all. Even for a bright quasar survey this is a concern regarding the photometric completeness.
We analyze the fraction of sources in the MQC without detections in the 2MASS and WISE bands necessary for the JKW2 color cut. 
More than $99.5\%$ of $m_{\rm{i}}<18.0$ quasars at $z{>}2.5$ in the SDSS matched MQC are detected with $\rm{SNR}>5$ in the W1 and W2 bands. 2MASS is more shallow but still allows for about ${>}80\%$ of $m_{\rm{i}}<18.0$ quasars at $z{>}2.5$ in the SDSS matched MQC to be detected in J and Ks.
The photometric completeness of all surveys will be taken account, when we calculate the overall survey completeness in Paper II.

\subsection{Galaxy Rejection using the Petrosian Radius} \label{sec_gal_rejection}

The SDSS survey offers the \texttt{type} parameter to differentiate between different source types. It allows to distinguish between point sources (\texttt{type}=6) and sources that are classified as extended (\texttt{type}=3).
However, we do not generally want to exclude quasar lenses in our selection and thus do not use this parameter. 
Instead we turn to a more quantitative measure, the petrosian radius.

At redshifts above $z{=}2.5$ even lensed quasars will be reasonably compact (\texttt{petroRad\_i}${<}2''$), if they are not strongly distorted, but might still be categorized as extended objects (SDSS flag \texttt{type}=3, e.g. the lensed quasar B 1422+231).

\begin{figure}[t]
 \includegraphics[width=0.5\textwidth]{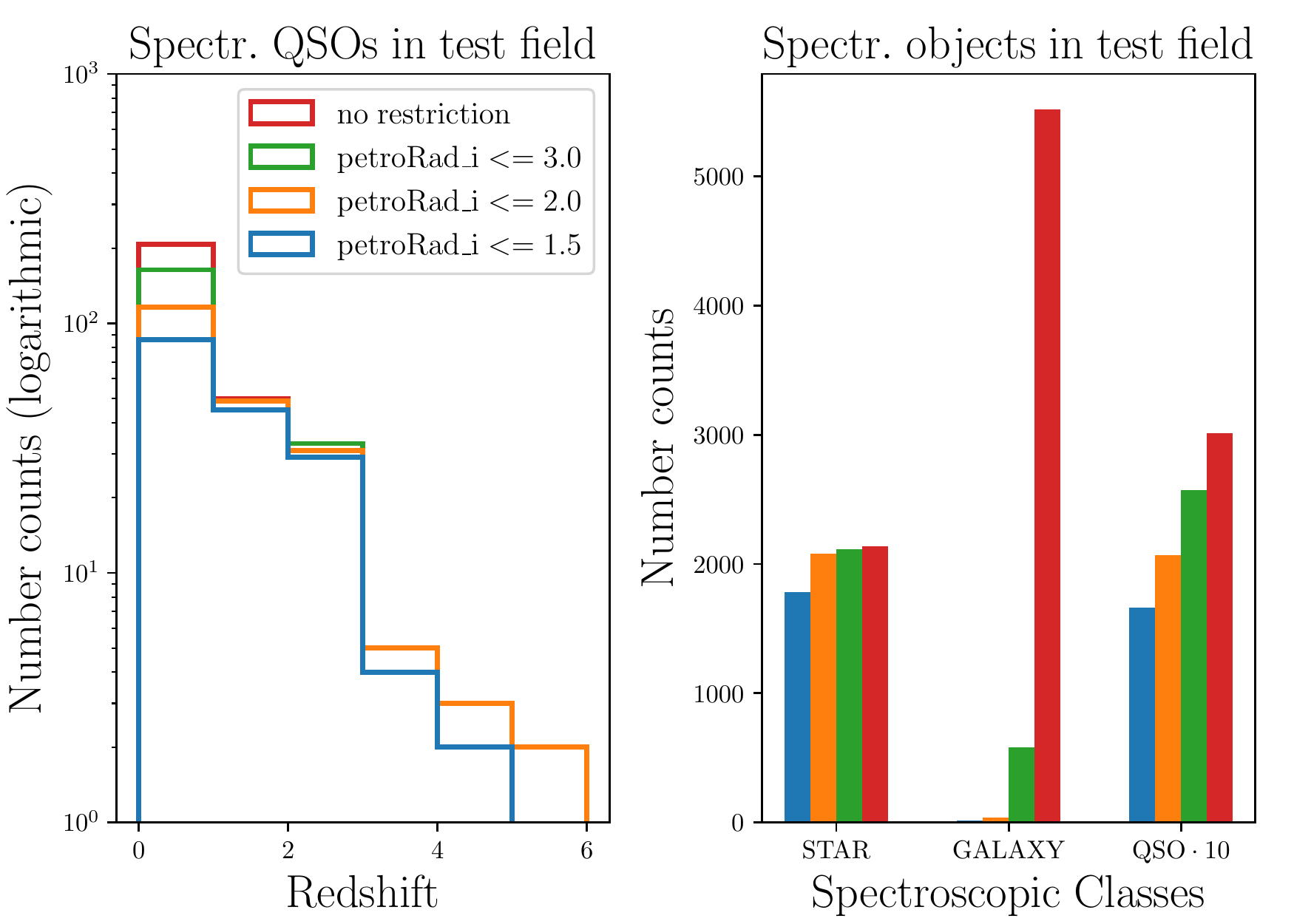}
 \caption{\textbf{Left:} The redshift distribution of the spectroscopically identified quasars in the test region as a function of cuts in petrosian radius in the SDSS i-band. \textbf{Right:} The total number of spectroscopically identified objects as a function of the limit on the petrosian radius in the SDSS i-band (red : unrestricted; green, orange, blue : \texttt{petroRad\_i} $\leq 3.0'', 2.0'', 1.5''$).}
 \label{fig_petrorad}
\end{figure}

In Figure\,\ref{fig_petrorad} we show that a petrosian radius of $2''$ in the SDSS i-band strongly reduces the number of galaxies in the total test region, while mainly reducing the number of quasars between $0 \le z\le 1.0$, where its host is  likely resolved in the SDSS photometry.
In the targeted redshift range of $2.5{\le}z{\le}4.0$ this cut on the petrosian radius in the SDSS i-band does not reject any quasars.

\begin{table*}[t]
\centering
\caption{Selection criterion on $70\,\rm{deg}^2$ test region:  Petrosian radius}
\label{tab_petrorad}
\begin{tabular}{ccccc}
\tableline
\tableline
Criterion& - & \texttt{petroRad\_i}$\le 3.0\,\rm{arcsec}$ & \texttt{petroRad\_i}$\le 2.0\,\rm{arcsec}$ & \texttt{petroRad\_i}$\le 1.5\,\rm{arcsec}$ \\
\tableline
Full sample  & 1,155,203 & 965,887 & 841,925 & 544,643 \\
Galaxies & 5,517 & 578 & 39 & 12 \\
Quasars & 301 & 257 & 207 & 166 \\
Stars & 2,138 & 2,112 & 2,081 & 1,781 \\
No spectral Id. & 1,147,247 & 962,940 & 839,598 & 542,684 \\
\tableline
\end{tabular}
\end{table*}

We summarize the effects of the cuts in petrosian radius on the $70\,\rm{deg}^2$ test region (Fig.\,\ref{fig_testfield_colorcut}) in Table\,\ref{tab_petrorad}.  
The limit on the petrosian radius reduces the number of spectroscopically identified quasars from 301 to 207 out of which 205 ($99\,\%$) are within the JKW2 color cut. 
As expected, the petrosian radius limit does not significantly alter the number of stars within the color cut.
Yet, it reduces the total number of spectroscopically identified galaxies that make the JKW2 color cut from 2706 to 15, and the total number of sources from 25470 to 1296. 
The majority of the rejected sources will be galaxies with a contribution from low redshift ($z{\le}1.0$) quasars.

\section{Photometric Redshift Estimation and Quasar-Star Classification}\label{sec_rf}

The JKW2 color cut is very efficient in rejecting stellar sources and the additional limitation on the petrosian radius excludes the majority of galaxies. 
However, we currently have no measure to exclude low redshift quasars, since the JKW2 color cut does select quasars independent of their redshift.  About $\sim95\%$ of the DR7 and DR12 quasars that make the JKW2 color cut are at lower redshifts. While this distribution is biased due to the selection of SDSS and BOSS quasar candidates, it still shows us that our sample will suffer from a large quantity of lower redshift quasars.
In addition, bright quasars ($m_{\rm{i}}\leq18.5$) would make up less than half of the sources selected by these two criteria, the other half still being stellar contaminants.

Hence, we decided to use the SDSS, 2MASS and WISE photometry to further estimate their photometric redshifts and also classify quasar candidates. 
Our methods of choice here are two supervised machine learning algorithms, support vector machines \citep{Vapnik1995, Burges1998, Vapnik1998} and random forests \citep{Breiman2001}. 

While the amount of galaxies included in the JKW2 color cut might seem non-negligible, the visual inspection of photometric cutouts efficiently reduces them further and our spectroscopic observations show that they are insignificant contaminants. Consequently we decided not to include them in the classification process.

We will first explain the construction of the training sets that both algorithms will rely on, then introduce the methods themselves and finally discuss the results.

\subsection{Training Sets}

\subsubsection{The Empirical Quasar Catalog}

In order to devise training sets for the classification and photometric redshift estimation, we combined the SDSS quasar catalogs with 2MASS All-Sky and WISE AllWISE photometry. 

We select all quasars from the SDSS DR7 and DR12 quasar catalogs (DR7Q and DR12Q, respectively) and matched them with SDSS DR13 (PhotoObjAll) sources in a $2''$ radius to obtain additional information about the sources. The resulting catalog is then matched to WISE AllWISE and 2MASS All-Sky photometry in a $1''$ radius to make sure that the matches are accurate. We exclude magnitude outliers ($2 {<} i$ and $i {>} 29$; $i{\equiv}$apparent i-band magnitude) from the catalog. The magnitudes are then corrected for galactic extinction ($m_{\rm{i}}{\equiv}$ extinction corrected apparent i-band magnitude).
Fluxes and flux errors (where possible) are then calculated from the extinction corrected magnitudes.
To ensure that the magnitude error distributions are free of outliers, objects with positive values in the SDSS flags \texttt{INTERP} and \texttt{DEBLEND\_AT\_EDGE} are excluded.

The resulting quasar catalog is termed the ``empirical'' quasar catalog and has a total of 215,087 objects with full SDSS photometry. An overview of the different sample sizes drawn from this catalog is given in Table\,\ref{tab_data_samples}.

The column ``photometry'' refers to the photometric information included in this training set. Not all sources in the full empirical quasar catalog have information in the 2MASS or WISE filter bands, because the 2MASS and WISE surveys do not always reach the same depth as the SDSS photometry. If we mandate values in the WISE W1 and W2 bands or the 2MASS bands and their corresponding errors the size of the training set decreases.

One additional question, that naturally arises, is: Can we find a bright population of quasars, that does not exist in the training set?
The distribution of the brightest quasars does not have inherently different spectral slopes. As a consequence their flux ratios (or colors) do not differ substantially from less brighter ones.
Since the main features used in the machine-learning methods are the flux ratios of adjacent photometric bands, bright quasars should be comprehensively selected without problems.

\begin{table}[h]
 \centering
 \caption{The different training sets for the photometric redshift estimation and quasar classification.}
 \label{tab_data_samples}
 \begin{tabular}{cccc}
 \tableline
 \tableline
  Data Set & Photometry & Constraints & Size \\
 \tableline
  Emp. QSOs & SDSS & - & 215,087 \\
  Emp. QSOs & SDSS & $m_i < 18.5$ & 12,408 \\
  Emp. QSOs & SDSS+W1W2 & - & 153,890 \\
  Emp. QSOs & SDSS+W1W2 & $m_i < 18.5$ & 12,388 \\
  Emp. QSOs & SDSS+2MASS+W1W2 & - & 4,815 \\
  Emp. QSOs & SDSS+2MASS+W1W2 & $m_i < 18.5$ & 4,021 \\
  Emp. QSOs & SDSS+2MASS+W1W2 & $m_i < 18.5$, JKW2 & 4,015 \\
  \tableline
  DR13 Stars & SDSS & - & 387,854 \\
  DR13 Stars & SDSS & $m_i < 18.5$ & 219,375 \\
  DR13 Stars & SDSS+W1W2 & - & 245,326 \\
  DR13 Stars & SDSS+W1W2 & $m_i < 18.5$ & 197,798 \\
  DR13 Stars & SDSS+2MASS+W1W2 & - & 174218 \\
  DR13 Stars & SDSS+2MASS+W1W2 & $m_i < 18.5$ & 159,211 \\
  DR13 Stars & SDSS+2MASS+W1W2 & $m_i < 18.5$, JKW2 & 209 \\
 \end{tabular}

\end{table}

\subsubsection{The Empirical Star Catalog}

To construct a catalog of stars, we have restricted our sample to only spectroscopically classified stars in the SDSS footprint. Some of those spectroscopic validations were a result of quasar selection and as such our star catalog may be biased towards stellar classes that can be confused with quasars. However, this works to our advantage as common quasar contaminants will be over-represented in the sample.

Using the Casjobs\footnote{\url{http://skyserver.sdss.org/casjobs/}} interface our query automatically added SDSS photometry with the 2MASS PSC and the WISE AllWISE catalogs where objects were pre-matched.
To ensure good quality photometry in the SDSS bands we have used the following quality flags: \texttt{zWarning}=0, \texttt{INTERP}=0, \texttt{BINNED1}!=0,
\textit{not} (\texttt{EDGE, NOPROFILE, NOTCHECKED, PSF\_FLUX\_INTERP, SATURATED}), \texttt{BAD\_COUNTS\_ERROR}=0, \texttt{DEBLEND\_NOPEAK}=0, \texttt{INTERP\_CENTER}=0, \texttt{COSMIC\_RAY}=0

The resulting catalog contains a total of 387,854 objects which have full SDSS photometric information. A detailed list of the size of different subsamples, which have more photometric information, is given in Table\,\ref{tab_data_samples}.

For the purpose of quasar-star classification we have summarized the number of stellar subclasses of SDSS DR13 (SpecObj) into the following stellar classes: O, OB, B, A, F, G, K, M, L, T, WD, CV and Carbon.
The number of object per class is shown in Table\,\ref{tab_star_spec_classes}.

\begin{table}
\centering
\caption{Spectroscopic star classes}
 \begin{tabular}{cr}
  \tableline
  \tableline
  Spectral class & \# of objects \\
  \tableline
  O & 547\\
  OB & 533\\
  B & 4638\\
  A & 47,714\\
  F & 158,233\\
  G & 29,953\\
  K & 60,233\\
  M & 64,691\\
  L & 1329\\
  T & 172\\
  WD & 15,782\\
  CV & 3363\\
  Carbon & 321\\
  \tableline
 \end{tabular}
\label{tab_star_spec_classes}
\end{table}

\subsection{Introduction to Random Forest Methods}

The random forest method \citep{Breiman2001} uses an ensemble of decision (or regression) trees to vote for the most popular class (value), regarding a classification (regression) problem. 
The method is a supervised machine learning method. Therefore it requires a training set to learn from. Furthermore random forests are non-parametric, inherently allow for multiple ($>2$) classes and avoid the problem of over-fitting.

Since random forests rely on decision (or regression) trees, we will shortly introduce their operation.

The features of the training set, which are fluxes and flux ratios for our purposes, generate the multi-dimensional input space for the classification or regression. A decision tree divides this input space into cuboid regions along the feature axis. Each of these regions contains one or more data points which determine the target class or target value for that cuboid.

This structure is build corresponding to a binary tree, where at each node a decision on one input feature is made on how to split the tree into two branches. This is called recursive binary partitioning. The decision is reached using a criterion that encourages the formation of regions, where the majority of the data points belong to only one class.

To predict the target class or target value for a new data point, the class or value of the cuboid region, that the data point falls into, is chosen.

Decision trees are invariant under scaling of the feature values and to the inclusion of irrelevant features. They also produce re-viewable models of the data, that are easy to visualize and inherently work with multiple ($>2$) classes. 

In the random forest method each decision tree is fit on a bootstrap sub-sample of equal size to the original input sample. 
During the construction of the decision tree a random sub-set of all available features is used to determine the best splits for all nodes. In our case the number of random features considered is taken to be square-root of the total number of features. A full decision tree is grown without pruning the tree during the construction.
The forest contains a large number of these trees ($>100$), each of which gives a classification (target value) for every object. The class with the most votes is then chosen to be the final class of the object. For random forest regression the mean of all target values is calculated to be the final regression value.

While one decision tree is prone to over-fitting the training sets, random forests overcome this problem through the averaging process of many randomized decision trees.

\citet{Breiman2003} first proposed this machine learning technique as an astronomical classification tool to find quasars. In astronomy the method has since been applied for classification of variable stars \citep{Dubath2011,Richards2011} and variable quasars \citep{Pichara2012}, photometric redshift estimation (mainly aimed at galaxies) \citep{Carliles2010,CarrascoKind2013} and quasar classification \citep{Carrasco2015}.

We are using the implementation of the random forest classifier and regressor provided by the \texttt{scikit-learn} \citep{scikit-learn} python library with many of the default parameters. 

For the construction of the binary tree we use the Gini impurity to determine the best split at each step.
While the original random forest method \citep{Breiman2001} lets each classifier vote for the final class, this implementation averages the probabilistic predictions of each classifier to find the final probabilities for each class.

We adopt the hyper-parameters that control the size of the decision trees (\texttt{min\_samples\_split}, \texttt{max\_depth}) as well as the size of the forest (\texttt{n\_estimators}) to find the best classification/regression model. The best-fit values for these hyper-parameters are found using a limited grid search.

\subsection{Introduction to Support Vector Machines}

Support vector machines (SVMs) \citep{Vapnik1995, Burges1998, Vapnik1998} offer a sparse kernel method for classification, regression and novelty detection. Similar to random forests SVMs belong to supervised machine learning methods and therefore rely on a well constructed training set. Fundamentally a two-class classifier, the extension of SVMs to multi-class ($>2$) classification is problematic.

In the case of classification the algorithm calculates a decision boundary (hyperplane) to divide the multi-dimensional feature space into two regions, according to the two classes. The decision boundary is constructed to maximize the smallest distance between itself and any of the data samples. In the end only a small subset of original data points is necessary to define the decision boundary. These data points are the support vectors. One strength of the SVM lies in this reduction of information to only a few data samples that define how to split the feature space. As a result the SVM is very fast in making predictions.

SVMs as kernel methods use an algorithm that allows for the use of a kernel function to implicitly transform the original feature space into an higher-dimensional feature space. In such an algorithm the input vector enters only in the form of its scalar product and is then substituted by the kernel. Therefore the coordinates in the higher-dimensional feature space don't have to be calculated explicitly, but only their inner product. 
The kernel allows for complex decision boundaries that are non-linear in the original feature space and allow for more complicated distributions of the two classes.

However, many data sets do not have fully separable classes even if non-linear kernels are used. 
As a solution the algorithm allows for training data to lie on the ``wrong'' side of the decision boundary. These data points are then misclassified but ignored by the algorithm. They are called slack variables and allow for a better generalization of the classification. In older formulations of the SVM algorithm the amount of slack variables are controlled with the parameter  C.

Similar to other regression problems SVM regression also seeks to minimize a regularized error function. This error function incorporates the previously introduced slack variables as well as an $\epsilon$-insensitive error function \citep{Vapnik1995}. 
Therefore data points within $\epsilon$ of the regression model have no associated error and the number of data points outside of this region is controlled using the slack variable parameter C.

SVMs have been widely used in astronomy for galaxy  \citep{Wadadekar2005, WangDan2008} and quasar \citep{Han2016} photometric redshift estimation and source classification \citep{Gao2008, HuertasCompany2008, Kim2012, Peng2012, Kurcz2016}. 

We use the \texttt{scikit-learn} \citep{scikit-learn} implementation of support vector regression for a comparison of the photometric redshift estimation with the random forest method above. We use the radial basis function kernel (\texttt{kernel='rbf'}), which adds the hyper-parameter \texttt{gamma}. 
In order to estimate the optimal hyper-parameters \texttt{C}, \texttt{epsilon} and \texttt{gamma}, we carry out limited grid searches.

Furthermore it is important to note that the features of the input and test vector need to be normalized because SVMs are not scale invariant.

\subsection{Photometric Redshift Estimation}

With the JKW2 color cut rejecting the majority of stars and the limit on the petrosian radius filtering out unwanted galaxies, lower redshift quasars become major contaminants for our survey targeted at $z{\geq}2.8$ quasars.
Hence, it is critical to estimate the quasar redshift and use this photometric redshift as a criterion for our candidate selection.
Instead of relying on optical color cuts \citep[e.g.][]{Richards2002}, we aim to utilize the full photometric information given with support vector machine regression (SVR) and random forest (RF) regression. 

The training sets for both algorithms are drawn from the empirical quasar catalog, which is based purely on the SDSS DR7 and DR12 quasar catalogs, as described above.

We use three different training sets build from the empirical quasar data, to test the effects of different feature sets and the size of the training sets on the regression outcome. 
The first (SDSS; Table\,\ref{tab_reg_results} rows 1 and 5) includes all sources with full SDSS photometry, the second (SDSS+W1W2; Table\,\ref{tab_reg_results} rows 3 and 7) includes W1 and W2 information in addition to full SDSS photometry, whereas the last (SDSS+W1W2 $m_{\rm{i}}<18.5$; Table\,\ref{tab_reg_results} rows 4 and 8) uses only sources with SDSS i-band magnitude brighter than $m_{\rm{i}}<18.5$ of the SDSS+W1W2 subset.
In general it is advisable to always include the largest amount of information possible. However because we want to evaluate the benefit of including the W1 and W2 bands in addition to SDSS photometry, we limit us to  SDSS photometry for a test case on the SDSS+W1W2 subset (Table\,\ref{tab_reg_results} rows 2 and 6).

We do not build a training set based on full SDSS, 2MASS and W1, W2 photometry, because the number of quasars with full 2MASS photometry is too small ($\sim1000$) to allow for sufficient training in the large feature space. 
In addition a cut on the i-band magnitude at $m_{\rm{i}}<18.5$ strongly reduces the number of higher redshift quasars in the training set. As a consequence the feature space is not sufficiently populated at higher redshifts and the method is biased against those quasars.

For training sets with only SDSS features we use the four adjacent flux ratios (u/g, g/r, r/i, i/z) from the five photometric bands and the SDSS i-band magnitude. When we include WISE photometry (SDSS+W1W2) we expand the flux ratios accordingly (+ z/W1, W1/W2) and add the W1-magnitude to the feature set.

For each subset we use a grid search on the hyper-parameters of the random forest and support vector Machine regression to determine the best regression model.
We calculate the regression results for all subsets using both machine learning methods to compare them to each other.

The grid for the random forest regression has the following hyper-parameters: \texttt{n\_estimators} = [50,100,200,300], \texttt{min\_samples\_split} = [2,3,4] and \texttt{max\_depth} = [15,20,25] (36 combinations). 

For the support vector Machine regression we use a grid of \texttt{C} = [10,1.0,0.1], \texttt{gamma} = [0.01,0.1,1.0],                \texttt{epsilon} = [0.1,0.2,0.3] (27 combinations).

We use 5-fold cross validation on the full training set and always train on 80\% of the full training set, while the remaining 20\% are used to test the regression.
Before we continue to discuss the results of the calculations, we will first introduce the typical regression metrics.

\begin{figure*}[ht]
 \includegraphics[width=\textwidth]{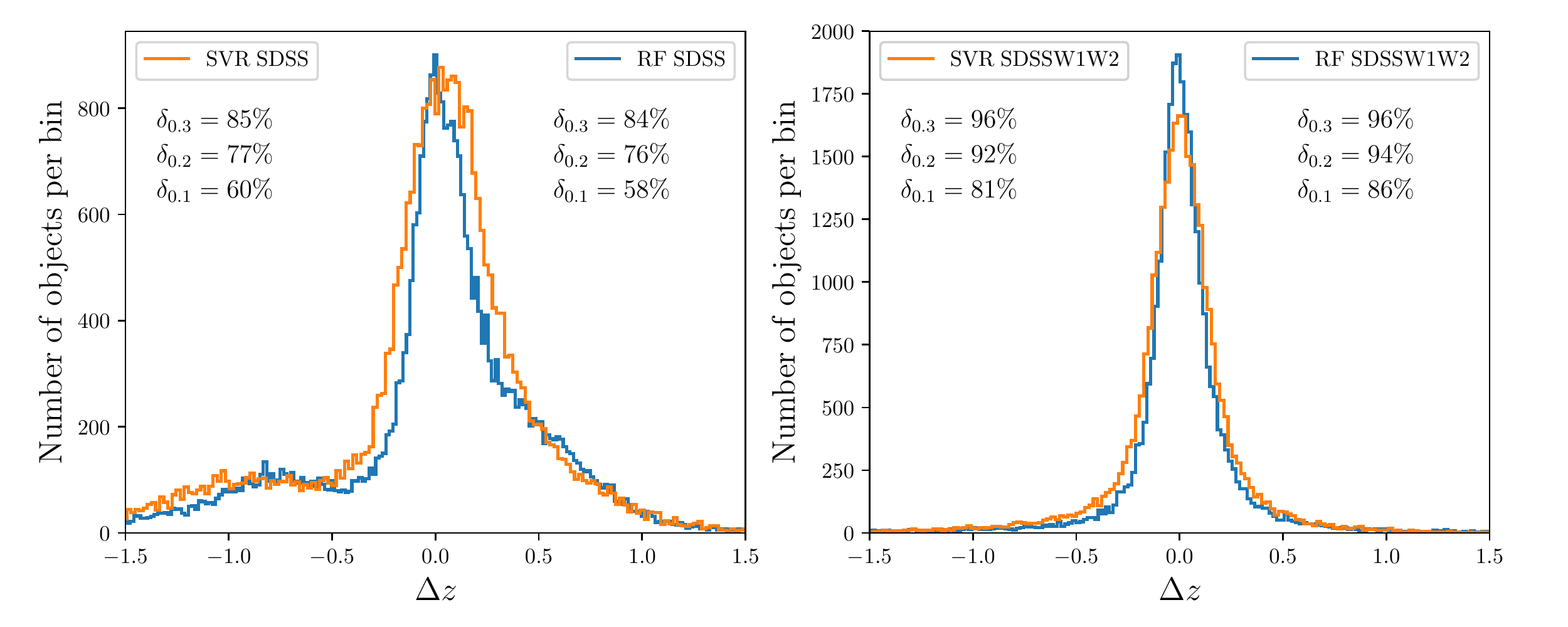}
 \caption{The distribution of the difference between the measured redshift of test quasars and their given regression value $\Delta z = z_{\rm{spec}}-z_{\rm{reg}}$. The left panel shows results from the SDSS feature set (rows 2 and 6 of Table\,\ref{tab_reg_results}) while the right panel includes the WISE W1 and W2 information (SDSS+W1W2, rows 3 and 7 of Table\,\ref{tab_reg_results}). The orange curves correspond to the support vector machine regression, whereas the blue curves are from the random forest method on the same training sets.}
 \label{fig_photoz_hist}
\end{figure*}

\begin{table*}[ht]
\centering
 \caption{Results of the Photometric Redshift estimation methods}
\begin{tabular}{ccccccccccc}
\tableline
\tableline
Data set & Training / Test size & Constraints & Features & Algorithm & $\delta_{0.3}$ & $\delta_{0.2}$ & $\delta_{0.1}$ & $\sigma$ & $R^2$ & \\
\tableline
DR7DR12Q & 172069 / 43018 & SDSS fl.r. & SDSS & RF & 0.87 & 0.81 & 0.65 & 0.483 & 0.654  &\\
 DR7DR12Q & 123112 / 30778 &SDSS+W1W2 fl.r. & SDSS & RF & 0.84 & 0.76 & 0.58 & 0.503 & 0.624 & \\
 DR7DR12Q  & 123112 / 30778 &SDSS+W1W2 fl.r. & SDSS+W1W2 & RF & 0.96 & 0.94 & 0.86 & 0.277& 0.884 & $\star$ \\
 DR7DR12Q & 9910 / 2478 &SDSS+W1W2 fl.r., $m_{\rm{i}}<18.5$ & SDSS+W1W2 & RF & 0.98 & 0.97 & 0.93 & 0.189 & 0.937 &\\
 \tableline
 DR7DR12Q & 172069 / 43018 &SDSS fl.r. & SDSS & SVR & 0.87 & 0.82 & 0.66 & 0.491 & 0.642 &\\
 DR7DR12Q & 123112 / 30778 &SDSS+W1W2 fl.r. & SDSS & SVR & 0.85 & 0.77 & 0.60 & 0.512 & 0.614 &\\
 DR7DR12Q & 123112 / 30778 &SDSS+W1W2 fl.r. & SDSS+W1W2 & SVR & 0.96 & 0.92 & 0.81 &0.291 & 0.873 &\\
 DR7DR12Q & 9910 / 2478 &SDSS+W1W2 fl.r., $m_{\rm{i}}<18.5$ & SDSS+W1W2 & SVR & 0.98 & 0.96 & 0.89 & 0.194 & 0.933 & \\
\tableline
\end{tabular}
\label{tab_reg_results}
\end{table*}

\subsubsection{Regression Metrics}
To evaluate the success of the photometric redshift estimation methods we use the standard $R^2$ regression score as well as the residual between photometric and spectroscopic redshift $\Delta z = z_{\rm{spec}}-z_{\rm{reg}}$.

The $R^2$ score, also called the coefficient of determination, gives a measure of the goodness of fit of a model. 
Let's assume that the true redshifts are denoted by $z_i$ and the predicted redshift values are denoted by $\hat{z}_i$. 
The $R^2$ score is then calculated from the total sum of squares $SS_{\rm{tot}}$ and the residual sum of squares $SS_{\rm{res}}$, where $\bar{z}$ is the mean redshift, according to:
\begin{align}
 SS_{\rm{tot}} &= \sum_i\left(z_i-\bar{z}\right)^2 \ , & SS_{\rm{res}} &= \sum_i \left(z_i-\hat{z}_i\right)^2 \ ,
\end{align}

\begin{equation}
 R^2 = 1- \frac{SS_{\rm{res}}}{SS_{\rm{tot}}} \ .
\end{equation} 
In the best case scenario the residual sum of squares will go to 0 and the $R^2$ score will reach 1. If the model always predicts the mean value of the training data $\bar{z}$, then it results in a $R^2$ score of 0. For bad models it is possible that the $R^2$ reaches negative values. 
We use the $R^2$ as a measure to compare models with another to find the model that represents the data best. Therefore we focus on the relative values of different models and do not interpret the absolute one.

In the literature the goodness of the photometric redshift estimation is often measured by the fraction of test quasars $N$ with absolute redshift residuals $|\Delta z| = \left| z_i - \hat{z}_i \right|$ smaller than a given residual threshold $e$ \citep{Bovy2012, Richards2015, Peters2015}.
\begin{equation}
 {\Delta z}_{e} = \frac{N(\left| z_i - \hat{z}_i \right| < e)}{N_{\rm{tot}}}
\end{equation}
Typical values chosen are $e=0.1,0.2,0.3$. 
However, in many cases the redshift normalized residuals are used instead:
\begin{equation}
 \delta_{e} = \frac{N(\left| z_i - \hat{z}_i \right| < e\cdot(1+z_i))}{N_{\rm{tot}}} \ .
\end{equation}

\subsubsection{Results}

The results of the regression calculations using RF regression and SVR are shown in Table\,\ref{tab_reg_results}.

The top half of the table details the results from the RF method, whereas the bottom half shows the corresponding results for the SVR. Both algorithms perform similarly well for all four training and feature sets used. If the WISE W1 and W2 photometry is included, the RF method performs slightly better. 

A comparison between the qualitative results of the two methods is given in Figure\,\ref{fig_photoz_hist}.
The left panel shows the results of the second and sixth row of Table\,\ref{tab_reg_results} while the right panel shows the third and seventh row of the same table.
The orange and blue curves are histograms of the redshift residuals $\Delta z$  for the SVR and RF regression, respectively. 
While the RF method shows a tighter distribution of the histogram around $\Delta z = 0$, the different colored curves show the same general qualitative behavior in both panels.

The main difference between both methods lies in the computing time. For the subset referenced with the $\star$ (Table\,\ref{tab_reg_results}) the computing time for RF regression amounted to $281\,\rm{s}$, whereas the same subset calculated with SVR took $4682\,\rm{s}$ for similarly good results.
This is an inherent drawback of the SVR method as the computing time scales with $\mathcal{O}(N^3)$, where the number of objects in the training set is $N$. 
For comparison the random forest method computing time scales with the number of trees in the Forest $T$ and the depth of each tree $D$ as $\mathcal{O}(T{\cdot}D)$.
Hence, for large training sets the random forest regression should always be preferred if both regression methods perform equally well.

If one compares the $R^2$ score of the second and third row of Table\,\ref{tab_reg_results} if becomes obvious that the inclusion of more photometric features clearly improves the regression results. While the $R^2$ score summarized the quantitative effect in one number, we show the qualitative differences in Figure\,\ref{fig_photoz_hist}. 
From the left to the right panel the distribution of test quasars tightens significantly around $\Delta z= 0$ and the second bump around $\Delta z \approx -0.8$ disappears.

The price is the reduction of the training sample from 172,069 quasars to 123,112. However, a comparison between the first and second row of Table\,\ref{tab_reg_results} shows that a larger training set improves the regression results if the same number of features are considered. This suggests, that an addition of training objects still leads to a better characterization of the target values in the feature space.

The best regression results are achieved on the SDSS+W1W2 training set with $m_{\rm{i}}<18.5$. Because we ignore the distribution of fainter quasars here, the photometric errors on this training set are smaller. As a result the regression achieves better results. 
However, we have to advise caution here. The $m_{\rm{i}}<18.5$ cut severely reduces the number of training and test objects. Because the redshift distribution of the empirical training set is dependent on the i-band magnitude of the quasars, it biases the training set against higher-redshift sources.

We choose the RF method with the SDSS+W1W2 training set and all available features for our photometric redshift estimation (marked with a $\star$ in Table\,\ref{tab_reg_results}).

\begin{figure}[t]
 \includegraphics[width=0.5\textwidth]{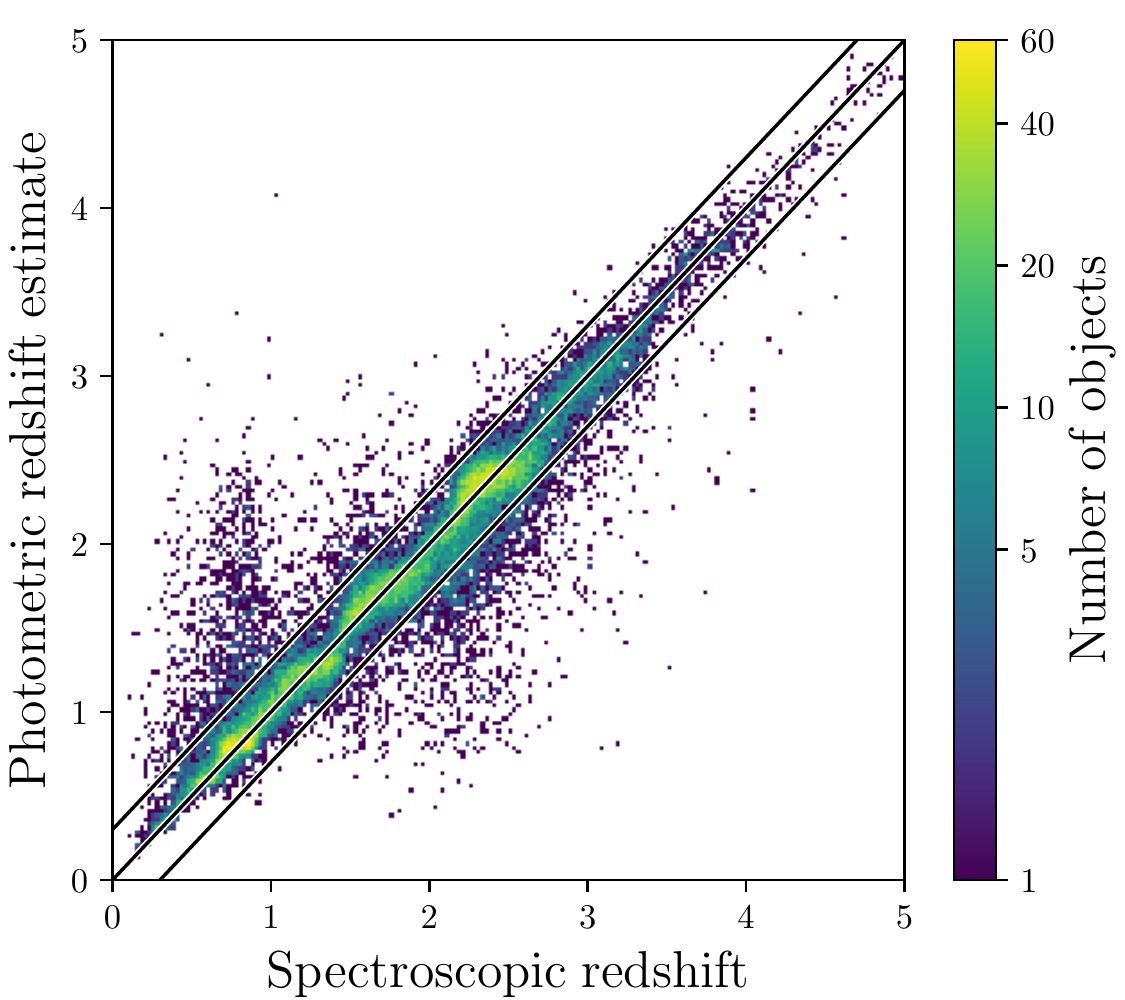}
 \caption{Photometric redshift  estimate of our test set calculated using the random forest method with training and feature set $\star$ against the measured spectroscopic redshift. The color bar shows the number of objects per rectangular bin. The three solid black lines illustrate the $\Delta z = 0$ diagonal and the $|\Delta z| =0.3$ region.}
 \label{fig_photoz_zz}
\end{figure}

Figure\,\ref{fig_photoz_zz} shows the density map of the random forest regression test results using this training set as a function of the measured spectroscopic redshift
The color of the map corresponds to the number of test objects per rectangular bin. 
The photometric redshift of the majority of test objects is well estimated. However a distribution of outliers still persists at spectroscopic redshifts of $z\approx 0.8, 1.6$ and $2.0-2.3$. 

\subsubsection{Comparison to other methods}
The difficulty in comparing different methods for redshift estimation to another is that the same training set should be used in all cases to create the model. Otherwise model comparisons, even using the same regression metrics, are not meaningful.

In a recent study on the photometric selection of quasars \citep{YangQian2017}, the authors incorporate asymmetries into a model of the relative flux distributions of quasars.
As part of their work they carry out a comparison \cite[see their Table\,1][]{YangQian2017} between their new method (Skew-t) and a range of frequently used algorithms in the literature (XDQSOz \citet{Bovy2012}, KDE \citet{Silverman1986}, CZR \citet{Weinstein2004}) based on the same photometric test and training sample. 
Their method achieves slightly better results than the other ones on the same training set based only on SDSS photometry.

While their training/test sample includes a range of high redshift quasars from the literature, the majority of the sample is also based on the DR7 and DR12 quasar catalogs.
Therefore we compare our results (Table\,\ref{tab_reg_results}) with theirs in Table\,\ref{tab_reg_comp}.
The RF and SVR are outperformed by the Skew-t method on the SDSS feature/training set (see $\delta z=0.1$), meanwhile the RF method performs equally well, once the WISE W1 and W2 bands are included.
Hence, random forests can keep up with other modern methods of photometric redshift estimation.

\begin{table}[h]
\centering
\caption{Comparison of the photo-z regression}
\label{tab_reg_comp}
\begin{tabular}{ccccc}
\tableline
 Method & Features/ Training set & $\delta_{0.2}$ & $\delta_{0.1}$ & Total data set size \\
\tableline
 Skew-t & SDSS & 0.82 & 0.75 &304,241\\
 RF & SDSS & 0.81 & 0.65 & 215,087 \\
 SVR & SDSS & 0.82 & 0.66 & 215,087 \\
 \tableline
 Skew-t & SDSS+W1W2 & 0.93 & 0.87 & 229,653 \\ 
 RF & SDSS+W1W2 & 0.94 & 0.86 & 153,890 \\
 SVR & SDSS+W1W2 & 0.92 & 0.81 & 153,890\\ 
 \tableline
 \end{tabular}
\end{table}

\subsection{Quasar-Star Classification}

To classify our quasar candidates as stars or quasars we only use the random forests, because the support vector machine algorithm is not suited for simultaneous classification with more than two classes.

The training set for the classification problem is build from the empirical quasar and star catalogs described above. 
Both catalogs are added to form the full empirical training set on which the random forest classification for different stellar spectral classes and quasar redshift classes will be trained.

However we limit ourselves to stars with spectral classes A,F,G,K,M, since these classes have enough objects for sufficient training and the main contaminants of quasars with redshifts $z=2-5$ are K and M stars. Most L and T dwarfs are too faint for our extremely luminous quasar survey and conflict only with even higher redshift quasars in color space. 
We have also excluded O and B stars because they are very rare and their blue colors are hardly confused with $z>2$ quasars.

Since quasar colors change considerably over redshift, we divide quasars into four redshift classes similar to \cite{Richards2015}. The classes are designed to split the quasars at redshifts, where dominating features change the u-band to g-band flux ratio. The redshift classes are ``vlowz'' with $0{<}z{\leq}1.5$, ``lowz'' with $1.5{<}z{\leq}2.2$, ``midz'' with $2.2{<}z{\leq}3.5$ and ``highz'' with $3.5{<}z$. At $z=1.5$ the Ly$\alpha$ emission line is just blueward of the u-band and the CIV emission line is still in the g-band. The second break at $z=2.2$ marks the point where the Ly$\alpha$ line is leaving the u-band and the last break at $z=3.5$ is marked by a strong flux decrease in the u-band as the Ly$\alpha$-forest absorbs flux blueward of the Ly$\alpha$ line.
We use these four quasar classes and five stellar classes to form the nine main labels for our classification problem. 
For evaluation purposes we introduce the binary labels ``STAR'' and ``QSO'', that encompass all stellar classes and all quasar classes, respectively. We will use them to calculate the completeness of the quasar selection against stars in general.

The RF algorithm is trained only on the photometric information. For classifications using only SDSS color space the features are the four adjacent flux ratios (u/g, g/r, r/i, i/z) from the five photometric bands and the SDSS i-band magnitude. When we include the 2MASS or WISE photometry we expand the flux ratios accordingly and add the J-band magnitude or the W1-magnitude to the feature set.

We investigate the effects of two magnitude limits in the SDSS i-band, $m_{\rm{i}}<21.5$ and $m_{\rm{i}}<18.5$, and the use of different sets of photometric features (SDSS, SDSS+W1W2 and SDSS+2MASS+W1W2) on the results of the fandom forest classification. 
The six subsets of the full empirical training set have different sizes, since not every object has always the full photometric information required. If information in one photometric band or its associated error is missing, we reject the source from the training set.

For each of the six subsets we find the best combination of the training hyper-parameters by calculating the classification results on a grid of \texttt{n\_estimators} =  [50,100,200,300], \texttt{min\_samples\_split} =  [2,3,4] and \texttt{max\_depth} = [15,20,25] (36 combinations). 
We use 5-fold cross validation and always train on 80\% of the full training set, while the remaining 20\% are used to test the classification.

\begin{table*}[t]
\caption{Results of the random forest classification on the full empirical training set}
\begin{tabular}{ccccccc}
 \tableline
 \tableline
 Training / Test size & Constraints & Features & p / r / F1 (highz) & p / r / F1 (QSO) & p / r / F1 (STAR)  & \\
 \tableline
  183108 / 45778 &  SDSS fl.r., $m_{\rm{i}}<18.5$ & SDSS & 0.92 / 0.71 / 0.80& 0.88 / 0.96 / 0.92 & 1.00 / 0.99 / 1.00 & \\
  167076 / 41770 &  SDSS+W1W2 fl.r., $m_{\rm{i}}<18.5$ & SDSS+W1W2 & 0.97 / 0.86 / 0.91& 0.91 / 1.00 / 0.95 & 1.00 / 0.99 / 1.00 & \\
  129934 / 32485 &  SDSS+2MASS+W1W2 fl.r., $m_{\rm{i}}<18.5$ & SDSS+2MASS+W1W2 & 0.88 / 0.88 / 0.88& 0.93 / 0.99 / 0.96 & 1.00 / 1.00 / 1.00  & \\
  442529 / 110634 &  SDSS fl.r., $m_{\rm{i}}<21.5$ & SDSS & 0.87 / 0.87 / 0.87 & 0.77 / 0.94 / 0.85 & 0.96 / 0.85 / 0.90 & \\
  313453 / 78364 &  SDSS+W1W2 fl.r., $m_{\rm{i}}<21.5$ & SDSS+W1W2 & 0.92 / 0.95 / 0.93 & 0.88 /1.00 /0.94 & 1.00 / 0.92 / 0.96& $\star$ \\
  141837 / 35460 &  SDSS+2MASS+W1W2 fl.r., $m_{\rm{i}}<21.5$ & SDSS+2MASS+W1W2 & 1.00 / 0.69 / 0.82 & 0.90 /0.99 /0.94 & 1.00 / 1.00 / 1.00 & \\
 \tableline
\end{tabular}
\label{tab_class_results}
\vspace{0.5cm}
\end{table*}
\subsubsection{Classification Metrics}

In order to measure the performance of the classification, we will introduce three standard classification metrics: The precision, the recall and the $F_1$-score \citep{Bishop2006}.

Precision (\textit{p}) is defined as the ratio of the true positives ($t_p$) to the sum of true and false positives ($t_p+f_p$). 
Usually in quasar selection one speaks of the purity/efficiency of the selection synonymous to the precision of the selection.

Whereas the recall (\textit{r}) is the ratio of true positives to the sum of true positives and false negatives ($t_p+f_n$). 
Regarding quasar selections the recall is equivalent to the completeness of the selection.

\begin{align}
 p &= \frac{t_p}{t_p+f_p} &  r &= \frac{t_p}{t_p+f_n}
\end{align}

The harmonic mean of precision and recall values is the traditional F-measure or balanced F-score. The 
$F_1$-score reaches its best value at 1 and the worst score at 0.

\begin{equation}
 F_1 = 2 \cdot \frac{\rm{precision}\cdot\rm{recall}}{\rm{precision}+\rm{recall}}
\end{equation}

For multi-class classification problems one can define a precision, recall and $F_1$ score for each class individually against all other classes.

Also an average precision, recall and $F_1$ score weighted by the number of occurrences in each true class, can provide an idea of how well the classifier generally works for a problem with multiple classes.

A helpful visualization for the results of classification problems is the confusion matrix $C$. Each entry $C_{i,j}$ is the number of objects known to be in class $i$, but predicted in class $j$. Therefore the entries $C_{i,i}$ show the true positives ($t_p$) for each class $i$. 
All other values in the row $i$ show the number of false negatives ($f_n$), objects predicted to belong to other classes, while they truly belong to class $i$. 
All other values in the column $i$ are the false positives ($f_p$), the values predicted to belong to class $i$, while they are truly belonging to other classes in the sample.

\subsubsection{Results}

\begin{figure}[t]
 \centering
 \includegraphics[width=0.5\textwidth]{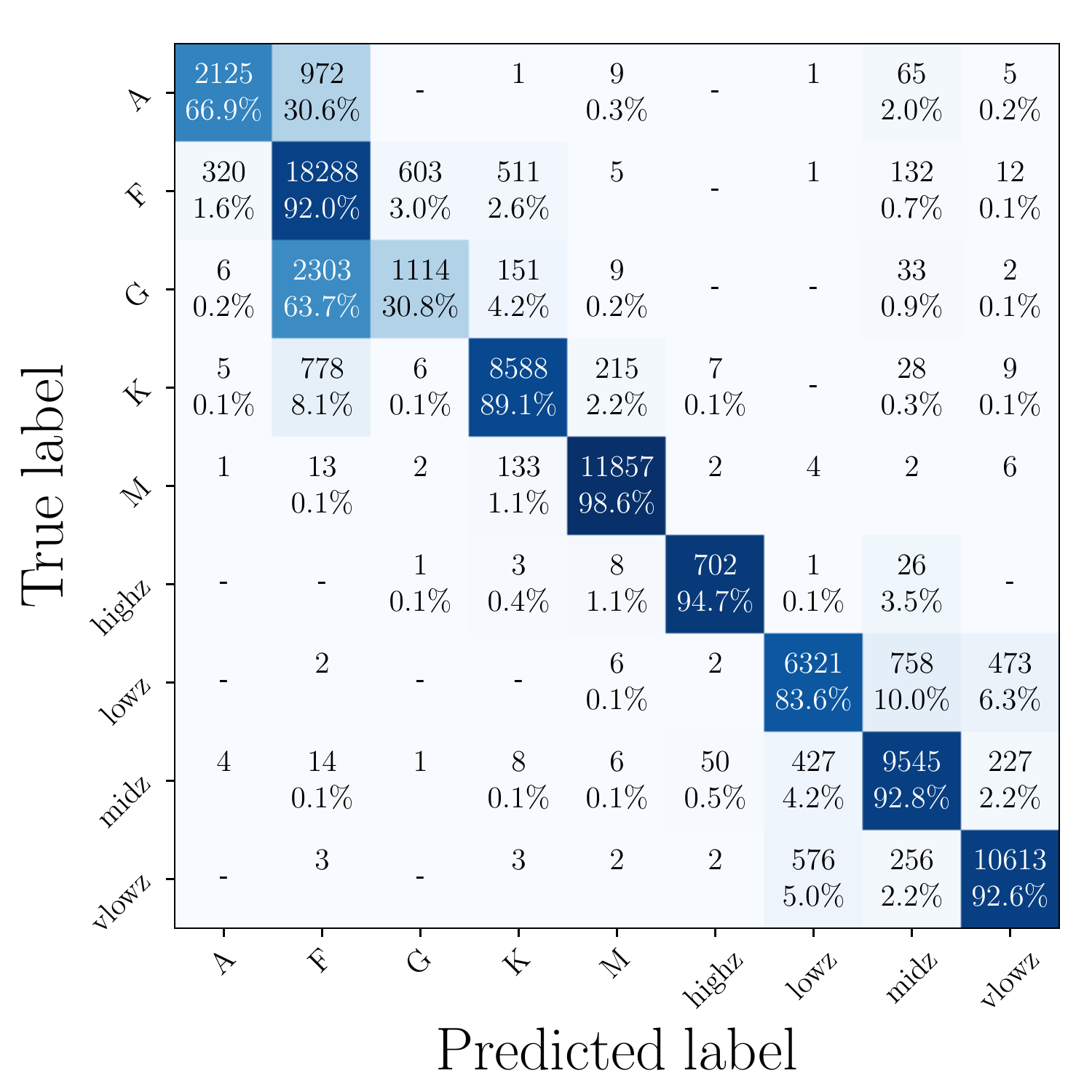}
 \caption{We show the confusion matrix for the classification using the W1 and W2 bands in addition to all SDSS photometry (subset SDSS+W1W2 $m_{\rm{i}}<21.5$). The true labels of the test set are on the row axis, whereas the columns refer to the predicted labels. The number of objects in each entry is displayed along with the percentage of that entry against the total number of objects with the same true label. Therefore all percentages in a row add up to 100\%. The color coding highlights the entries per row with the majority of objects.}
 \label{fig_conf_matrix}
\end{figure}

We present the results of the random forest classifications in Table\,\ref{tab_class_results}.
It shows the six different subsets of the full empirical training set along with their respective constraints and features used. 
We show the results of the classification as the precision (\textit{p}), recall (\textit{r}) and $F_1$-measures for the ``highz'' quasar class as well as the grouped classes of all quasars (''QSO``) and stars (''STAR``).

As expected, it is evident from Table\,\ref{tab_class_results}, that the inclusion of more photometric features always leads to better classification results at the prize of the training sample size.

The subsets with the same amount of features but different limits on the SDSS i-band show that the brighter samples ($m_{\rm{i}}<18.5$), the ones with the qualitatively better photometry, also show more accurate classifications. 
Again, the training sets for those three subsets are much smaller and will not be able to populate the entire feature space as well as a fainter limit of $m_{\rm{i}}<21.5$ would allow.

Since the number of higher redshift quasars is a strong function of the i-band magnitude any limitation of the training set in this regard will bias the classification of those objects.
For example the training set with full SDSS+2MASS+W1W2 photometry and $m_{\rm{i}}<18.5$ ($m_{\rm{i}}<21.5$) has only 38 (42) ''highz``($z>3.5$) training objects and 8 (13) test objects.
This demonstrates that these classification results cannot be fully trusted, because of the low number of objects that the precision, recall and $F_1$-score are based upon.

\begin{table}[t]
\centering
\normalsize
\caption{The binary label classification matrices of the SDSS+W1W2 subset ($m_{\rm{i}}<21.5$) with all features (top) and only the SDSS features (bottom)}
\begin{tabular}{p{2cm}|p{2cm} p{2cm}}
 \tableline
 \tableline
  \textbf{SDSS+W1W2} & pred. STAR & pred. QSO\\
 \tableline
 STAR & 48015 & 309\\
 QSO & 61 & 29979 \\
\end{tabular}
 \begin{tabular}{p{2cm}|p{2cm} p{2cm}}
\tableline
 \tableline
  \textbf{SDSS} & pred. STAR & pred. QSO\\
 \tableline
 STAR & 47747 & 577\\
 QSO & 333 & 29707 \\
 \tableline
\end{tabular}

\label{tab_conf_mat}
\end{table}

Based on these insights we adopt the SDSS+W1W2 subset with $m_{\rm{i}}<21.5$ as the best training and feature set for the quasar classification in our ELQS quasar selection (see $\star$ in Table\,\ref{tab_class_results}). 
It achieves the highest completeness (recall) of ''highz`` quasars and ''QSO`` in general of all subsets in Table\,\ref{tab_class_results} with the hyper-parameters set to \texttt{n\_estimators} = 300, \texttt{min\_samples\_split} =  3 and \texttt{max\_depth} = 25.

For this particular training and feature set we show the full confusion matrix in Figure\,\ref{fig_conf_matrix}. The rows correspond to the true class of the object, while the columns show the predicted labels. The values on the diagonal are the correctly classified objects. 
Each entry in the matrix shows the total number of objects and the percentage of the objects in that entry with respect to the true class (the full row). The entries are colored coded based on this percentage with a darker blue color corresponding to a higher percentage.

While the stellar classification encounters difficulties for the A,F and G stars the later spectral types and the quasars are well classified with diagonal entries above 80\%. 
Only a very small percentage of stars are classified as quasars (top right corner) with the majority of stellar contaminants falling into the ''midz`` quasar class ($2.2{<}z{<}3.5$). This is the redshift range in which the quasar distribution overlaps with the stellar locus the most in optical color space.

Conversely the bottom left corner shows quasars classified as stellar sources. Again the majority of these objects fall into the ''midz`` quasar class.

We can simplify this confusion matrix by grouping all stellar spectral classes and all quasar classes to the binary classes ''QSO`` and ''STAR``. The binary classification results are shown in Table\,\ref{tab_conf_mat}. Here the top part summarizes the results of the confusion matrix in Figure\,\ref{fig_conf_matrix}. 
In the bottom part we have used the same training set but reduced the feature set to only include the SDSS photometry. As a result the off-diagonal values are much higher, reflecting a larger fraction of stellar contaminants and a lower quasar completeness.
This demonstrates how the information gained by including the WISE W1 and W2 bands enhances the performance of the classifications.

\subsection{On the prospect of using only Random Forests for Quasar Selection}

We have demonstrated that random forests efficiently classify quasars and estimate their photometric redshifts using the SDSS optical bands in concert with WISE W1 and W2 photometry. 
In this context, we need to ask whether the JKW2 color cut is really necessary for our quasar candidate selection.

To evaluate this questions we assume the RF classification contamination as shown in Table\,\ref{tab_conf_mat} and apply it to the spectroscopically unidentified sources in the $70\,\rm{deg}^2$ test regions (Table\,\ref{tab_colorcut}). In this case the number of stellar contaminant would at least rise from $\sim1000$, which make the color cut, to $\sim1700$. 
This illustrates how important the information of the 2MASS J and K bands is to reject stars. 

Unfortunately only a small fraction of all known quasars and stars in the training sets have well detected 2MASS J and K photometry. The resulting training samples are too small to properly populate the multi-dimensional feature space to allow for sufficient training of the RF model. 
In addition the quasar training set would be strongly biased against higher redshift ($z\gtrsim3$) objects, which have less bright apparent magnitudes and are therefore less likely to be detected by 2MASS.
Even though we are focusing on the brightest quasars, it is important for the redshift range above $z\approx3$ to be well populated in the training set to achieve reliable results in the classification and regression.

As long as J and K band photometry is limited to the brightest objects in the training sets, the JKW2 color cut will play an important role in rejecting stellar sources.

\section{The ELQS Quasar Candidate Catalog}\label{sec_qso_catalog}

\subsection{Area coverage of the ELQS Survey}
The  ELQS survey includes all SDSS photometry with galactic latitudes $\rm{b}{<}-20$ or $\rm{b}{>}30$. To estimate the effective area of the ELQS Survey, we are using the Hierarchical Equal Area isoLatitude Pixelization \citet[HEALPix][]{Gorski2005}. The process of the calculation and the general parameters used are identical to the description in \citet{Jiang2016}. 

The effective area of the full ELQS survey is $11,838.5\pm20.1\,\rm{deg}^2$, with a contribution of $7,601.2\pm7.2\,\rm{deg}^2$ from the spring ($90\,\rm{deg} {<} \rm{RA} {<} 270\,\rm{deg}$) sky and a contribution of $4,237.3\pm12.9\,\rm{deg}^2$ from the fall ($\rm{RA} {>} 270\,\rm{deg}$ and $\rm{RA} {<} 90\,\rm{deg}$) sky.

\subsection{Construction of the Candidate Catalog}

With all tools at hand, we begin the construction of the ELQS quasar candidate catalog in the SDSS footprint. An overview is given in Fig.\,\ref{fig_selection_flowchart}.
The general source selection is based on the WISE AllWISE catalog matched with photometry from the 2MASS PSC. Both surveys are all-sky and therefore include the galactic plane, where the source density of stars is extremely high and leads to confusion in the near- and infrared surveys. 
Therefore we restrict the quasar candidates to larger galactic latitudes and only include sources with either $\rm{b}{<}-20$ or $\rm{b}{>}30$. We also require the WISE W1 and W2 photometry to have a $\rm{SNR}>5$ and positive $J$ band magnitudes to exist for all objects. All sources that pass these criteria and obey the JKW2 color cut  $ \rm{K}-\rm{W2} \ge 1.8 - 0.848 \cdot \left(\rm{J}-\rm{K} \right)$ are selected in our WISE-2MASS-allsky catalog. It comprises a total of 3,376,354 sources.

We then proceed to match all of these sources to the SDSS DR13 (PhotoPrimary) catalog in a $3.96''$ aperture. We do not reject any objects based on their photometric flags. None of the \texttt{fatal} or \texttt{non-fatal} flags of \citet{Richards2002} are evaluated in our selection. We rather inspect the images of all quasar candidates ($\sim400$ objects) in the very end to be as complete as possible. The matched SDSS-WISE-2MASS catalog has a total of 1,690,813 objects.

In the next step we apply the criterion on the petrosian radius (\texttt{petroRad\_i} $\le 2.0$; see Section\,\ref{sec_gal_rejection}) to reject the majority of galaxis. 
We also require all sources to satisfy the i-band magnitude cut of $m_{\rm{i}}<18.5$. 

For the remaining candidates we calculate photometric redshifts and evaluate their quasar probabilities using RF regression and classification (see Section\,\ref{sec_rf}).
This demands that all sources have quantified photometric errors for all SDSS and WISE W1 and W2 photometry. 
The classification calculates the most probable class of the objects (\texttt{rm\_emp\_mult\_class\_pred}), the general quasar or star class (\texttt{rf\_emp\_bin\_class\_pred}) and the total probability of the object to belong to the quasar class (\texttt{rf\_emp\_qso\_prob}). The regression delivers the best estimate for the photometric redshift (\texttt{rf\_emp\_photoz}).

With this information at hand, all objects that obey the photometric redshift cut of $z_{\rm{reg}} > 2.5$ and are generally classified as quasars
(\texttt{rf\_emp\_bin\_class\_pred}=QSO) or fall into the high redshift inclusion boxes defined in \citet{Richards2002} are considered quasar candidates.

In addition to these \textit{primary} candidates, we  allow all objects that pass the photometric redshift cut and also have a probability of $>30\%$ to belong to the quasar class (\texttt{rf\_emp\_qso\_prob}$\geq0.3$) to be included as \textit{additional} candidates.
This results in a candidate catalog of 2253/1735  total/primary  sources out of which 920/876 are known quasars from the literature and 1333/859 are valid candidates for spectroscopic follow-up.
From this sample we prioritize bright high-redshift objects with $z_{\rm{reg}} \geq 2.8$ and $m_{\rm{i}}\leq 18.0$, that will make up the ELQS spectroscopic survey. These criteria leave 742/594 total/primary candidates out of which 341/327 are known and 401/267 need to be followed up.

In the last stages every candidate's photometry will be inspected before observation to check for image defects or obviously extended sources. Of the total/primary sample roughly 59\%/69\%  are good candidates, 10\%/8\% are blended with other sources in WISE, 9\%/6\% are extended and 22\%/17\% have bad image quality in at least one of the bands or are contaminated by bright sources close by.
The final ELQS quasar candidate catalog includes a total of 237 targets out of which 184 are primary candidates.

\section{Conclusion}\label{sec_conclusion}

In this paper we show that the SDSS and BOSS quasar surveys have systematically missed bright quasars at redshifts $z\sim3$ and above.
This is mainly due to stellar contamination at redshifts, where quasars overlap in optical color-space with the stellar locus, and the surveys' incomplete spectroscopic observations in the fall sky ($\rm{RA} {>} 270\,\rm{deg}$ and $\rm{RA} {<} 90\,\rm{deg}$) of the SDSS footprint.

We have developed a more inclusive quasar selection algorithm that is based on a near-infrared/infrared color criterion with high quasar completeness. Further inclusion of the SDSS optical photometry allows for galaxy rejection, classification of sources in stellar spectral types and quasar redshift classes and photometric redshift estimation. 
The latter tasks are accomplished using the random forest machine-learning method on a training sample of SDSS DR13 spectroscopic stars and quasars from the SDSS DR7 and DR12 quasar catalogs.

While the near-infrared/infrared color criterion is very effective, the limiting magnitude of the 2MASS survey only allows to use it at the brightest end of the quasar distribution. This limits the use of this criterion, as even our first estimates, taking the combined SDSS DR7 and DR12 quasar catalogs as a basis, show that we only reach $80\%$ photometric completeness for $m_{\rm{i}}<18.0$ quasars at $z{>}2.5$.
Future infrared surveys (e.g. EUCLID) with deeper photometry will be able to exploit this color cut to create a fainter, highly complete quasar sample.

The total/primary high-priority quasar candidate catalog comprises 237/184 objects with $z_{\rm{reg}} \geq 2.8$ and $m_{\rm{i}}<18.0$. Observations taken up to August 2017, have successfully identified a total of 67 quasars at $z\geq2.8$ in both total and primary candidate samples. We estimate the efficiency of our quasar selection on the spectroscopically completed ELQS spring sky sample. It includes 340  primary candidates of which 39 are newly identified quasars as part of ELQS and 231 are known quasars in the literature. The remaining primary candidates were either low-redshift quasars (36) or identified not to be quasars (35) by our survey. This results in an efficiency of roughly $\sim79\%$.
The efficiency predicted by the RF classification reaches $80\%-90\%$ for "midz" and "highz" quasars. The reason for our somewhat lower efficiency value is likely found in our loose quality criteria on the SDSS, WISE, and 2MASS
photometry. We do not use any of the standard SDSS, 2MASS or WISE quality flags and only rely on $SNR\geq5$ in the WISE W1 and W2 bands and the quality requirements of the 2MASS PSC.

In a forthcoming publication we will present the spectroscopic observations of the completed spring sky footprint of SDSS, along with a discussion on the full completeness of the ELQS survey and a first estimation of the bright end quasar luminosity function.
With the conclusion of the survey, a final publication will calculate the ELQS quasar luminosity over the full survey footprint and discuss the implications for the evolution of the brightest quasars and thus the most massive black holes.

\subsection*{Acknowledgements}

JTS, XF and IDM acknowledge support from the NSF grant AST 15-15115.
QY, JW, and LJ acknowledge support from the National Key R\&D Program of China (2016YFA0400703)
and from the National Science Foundation of China (grant 11533001).

This publication makes use of data products from the Two Micron All Sky Survey, which is a joint project of the University of Massachusetts and the Infrared Processing and Analysis Center/California Institute of Technology, funded by the National Aeronautics and Space Administration and the National Science Foundation.
This publication makes use of data products from the Wide-field Infrared Survey Explorer, which is a joint project of the University of California, Los Angeles, and the Jet Propulsion Laboratory/California Institute of Technology, funded by the National Aeronautics and Space Administration.

Funding for the Sloan Digital Sky Survey IV has been provided by the Alfred P. Sloan Foundation, the U.S. Department of Energy Office of Science, and the Participating Institutions. SDSS acknowledges support and resources from the Center for High-Performance Computing at the University of Utah. The SDSS web site is \url{www.sdss.org}.

SDSS is managed by the Astrophysical Research Consortium for the Participating Institutions of the SDSS Collaboration including the Brazilian Participation Group, the Carnegie Institution for Science, Carnegie Mellon University, the Chilean Participation Group, the French Participation Group, Harvard-Smithsonian Center for Astrophysics, Instituto de Astrofísica de Canarias, The Johns Hopkins University, Kavli Institute for the Physics and Mathematics of the Universe (IPMU) / University of Tokyo, Lawrence Berkeley National Laboratory, Leibniz Institut f\"ur Astrophysik Potsdam (AIP), Max-Planck-Institut f\"ur Astronomie (MPIA Heidelberg), Max-Planck-Institut f\"ur Astrophysik (MPA Garching), Max-Planck-Institut f\"ur Extraterrestrische Physik (MPE), National Astronomical Observatories of China, New Mexico State University, New York University, University of Notre Dame, Observatório Nacional / MCTI, The Ohio State University, Pennsylvania State University, Shanghai Astronomical Observatory, United Kingdom Participation Group, Universidad Nacional Autónoma de México, University of Arizona, University of Colorado Boulder, University of Oxford, University of Portsmouth, University of Utah, University of Virginia, University of Washington, University of Wisconsin, Vanderbilt University, and Yale University.

This research made use of Astropy, a community-developed core Python package for Astronomy (Astropy Collaboration, 2013, \url{http://www.astropy.org}).

\bibliographystyle{apj}

\end{document}